\documentclass[11pt]{article}
\usepackage[utf8]{inputenc}
\usepackage[english]{babel}
\usepackage{amsmath,amsfonts,amsthm,graphicx,float}
\usepackage{natbib}

\DeclareMathAlphabet{\mathpzc}{OT1}{pzc}{m}{it}

\usepackage{algorithm}

\usepackage[noend]{algpseudocode}

\def\bm{{\mathbf m}}

\def\b1{{\mathbf 1}}

\usepackage[font=small]{caption}

\usepackage{titlesec}

\titleformat*{\section}{\normalfont\fontsize{14}{17}\bfseries}
\titleformat*{\subsection}{\normalfont\fontsize{12}{15}\bfseries}

\usepackage{bm}
\usepackage{enumerate}
\providecommand{\abs}[1]{\lvert#1\rvert}
\providecommand{\norm}[1]{\lVert#1\rVert}

\usepackage{soul}
\usepackage[usenames, dvipsnames]{color}

\date{}

\begin{document}

\title{\LARGE Goodness-of-fit tests for parametric regression models with circular response}\normalsize
\author{Andrea Meil\'an-Vila \\
Universidade da Coru\~{n}a\thanks{%
Research group MODES, CITIC, Department of Mathematics, Faculty of Computer Science, Universidade da Coru\~na, Campus de Elvi\~na s/n, 15071,
A Coru\~na, Spain}
\and %
Mario Francisco-Fern\'andez\\
Universidade da Coru\~{n}a\footnotemark[1]
\and
Rosa M. Crujeiras \\
Universidade de Santiago de Compostela\thanks{Department of Statistics, Mathematical Analysis and Optimization, Faculty of Mathematics, Universidade de Santiago de Compostela, R\'ua Lope G\'omez de Marzoa s/n,
	15782, Santiago de Compostela, Spain}}
\maketitle


\begin{abstract}
Testing procedures for assessing a parametric {regression model with circular response and $\mathbb{R}^d$-valued} {covariate} are proposed and analyzed in this work{ both for} independent and  for spatially correlated data. The {test} statistics are  based on a circular distance comparing a ({non-smoothed} or smoothed) parametric {circular estimator} and a nonparametric {one}. {Properly designed bootstrap procedures for calibrating the {tests} in practice are also presented}. Finite sample performance of the {tests}{ in different scenarios with independent and spatially correlated samples,} is analyzed {by simulations}.

\end{abstract}	


\section{Introduction}\label{sec:gof_circular_int}
In many scientific fields, such as oceanography, meteorology or biology, data  are angular measurements (points in the unit circle {of a circular variable}), which are  accompanied by {auxiliary} observations of other Euclidean random variables. The joint behavior of these circular and Euclidean variables can be analyzed by considering a regression model, allowing at the same time to explain the possible relation between the variables and also to make predictions on the variable of interest. Parametric regression estimators for linear-circular models (circular {response} and Euclidean covariates) {with independent data} have been studied by \citet{fisher1992regression}, \citet{presnell1998projected}, and \citet{kim2017multivariate}, among others. In the presence of spatial correlation,  \citet{jona2012spatial}, \citet{wang2014modeling}, \citet{lagona2015hidden} and \citet{mastrantonio2016wrapped}{ for instance,} employed parametric methods to model {circular spatial processes}. Alternatively, nonparametric regression approaches can be used to deal with these inference problems. For this purpose, kernel-type estimators of the regression function in a model with a circular response and a $\mathbb{R}^d$-valued {covariate have been introduced by \citet{meilan2020a,meilan2019c}}.  Notice that if the bandwidth matrix is appropriately chosen, these methods provide more flexible estimators than those obtained using parametric approaches, {avoiding misspecification problems}. However, if {a} parametric regression model {is assumed and it} holds, parametric methods usually {provide estimators which are more efficient and easier to interpret.} In this context, goodness-of-fit tests can be designed, providing a tool for assessing a general class of parametric linear-circular regression models. 

There is a substantial literature on testing  parametric regression models involving Euclidean data, including \citet{kozek1991nonparametric}, \citet{hardle1993comparing}, \citet{gonzalez1995testing},  \citet{biedermann2000testing}, \citet{park2015using}, {\citet{meilan2020b}, and \citet{meilan2019goodness}}, among others. {See \citet{gonzalez2013updated} for a review}. The previous testing procedures  are based on measuring differences between {a suitable parametric estimator under the null hypothesis and a nonparametric one}. Specifically,  $L_2$-norm or supremum-norm tests, among others, can be employed for regression models with Euclidean responses and covariates. In the context of regression models with directional response and directional or Euclidean explanatory variables, 
the literature on goodness-of-fit tests is relatively scarce. {In this setting,} \citet{deschepper2008tests} proposed an exploratory tool and a lack-of-fit test for circular-linear
regression models (Euclidean response and {circular} covariates). The same problem was studied by \citet{garcia2016testing}, {using nonparametric methods}. The authors proposed a testing procedure
based on the weighted squared distance between a smooth and a parametric regression
estimator, where the smooth regression estimator was obtained by a projected local {regression} on the sphere. However, the problem of assessing a certain class of parametric linear-circular regression models{, that is, for a regression model with circular response and $\mathbb R^d$-valued covariates} (up to the authors knowledge) has not been considered in the statistical literature yet, neither for independent nor for spatially dependent observations.

In this work, new approaches for testing a linear-circular parametric regression model (circular {response} and  {$\mathbb R^d$-valued {covariate}}) are proposed and analyzed{ both for independent and spatially correlated errors.}  The test statistics employed  in these procedures are based on a comparison between  a {(non-smoothed or smoothed)} parametric fit  {under the null hypothesis} and a nonparametric estimator of the circular regression function. More specifically, two different test statistics are considered. In the first one, the parametric estimator of the regression function {under the null hypothesis} is directly used, while in the second one, a smooth version of this estimator is employed. Notice that, in this framework, a suitable measure of \emph{circular distance} must be employed \citep[see][Section 1.3.2]{jammalamadaka2001topics}. The null hypothesis that the regression function {belongs to a certain parametric family} if the distance between both fits exceeds a certain threshold.  To perform the parametric estimation, a circular analog to
least squares regression   is used \citep[see][]{fisher1992regression,lund1999least}.  For the nonparametric alternative, {kernel-type} regression  estimators \citep{meilan2020a,meilan2019c} {are considered.}

{For the application in practice of the proposals, the test statistics should be accompanied by a calibration procedure. In this {case}, this is not based on the asymptotic distribution, given that the convergence to the limit distribution under the null hypothesis {will presumably be} too slow. Different bootstrap methods are designed and their performance is analyzed and compared in empirical experiments.}  For independent data, {standard} resampling procedures adapted to the context of regression models with circular {response} are used: a parametric circular residual bootstrap (PCB) and a nonparametric circular residual bootstrap (NPCB). The PCB approach consists in using {the residuals obtained from the parametric fit in the bootstrap algorithm}. If the circular regression function belongs to the parametric family considered in the null hypothesis, then the residuals will tend to be quite {similar} to the theoretical errors and, therefore, it is expected that the  PCB method has a good performance. Following the proposal by \citet{gonzalez1993testing}, the NPCB method aims to increase the power of the test and, for this purpose, the residuals {obtained from the nonparametric fit are the ones {employed} in the bootstrap procedure}. The previous resampling procedures (PCB and NPCB) for independent data {must be properly adapted for handling} spatial correlation. Two specific procedures for test calibration which take the {spatial correlation into account are also introduced}: a parametric spatial circular residual bootstrap (PSCB) and a nonparametric spatial circular  residual bootstrap  (NPSCB).  {Similarly to the PCB, but now for spatially correlated errors, the PSCB considers the residuals obtained from the parametric fit under the null hypothesis}. The relevant difference between PCB and PSCB is that, in order to mimic the dependence structure of the errors, a spatial circular process  is fitted to the residuals in PSCB. Samples coming from the fitted process are employed in the bootstrap algorithm.  The steps followed in NPSCB are similar at those employed in PSCB, but the residuals are obtained from the nonparametric regression estimator.

This paper is organized as follows. Section \ref{sec:gof_circular_ind} is devoted to present some ideas of goodness-of-fit tests for circular regression models. The parametric and nonparametric circular regression estimators employed in the test statistics are  presented in Sections \ref{sec:gof_circular_estimation_par} and \ref{sec:gof_circular_estimation_npar}, respectively. Section \ref{sec:gof_circular_testing} introduces the testing problem and the proposed test statistics. A  description of the calibration algorithms considered is given in Section \ref{sec:gof_circular_ind_cal}. Section \ref{sec:gof_circular_ind_simu} contains a simulation study for assessing the performance of the tests when using the PCB and NPCB resampling approaches to approximate the sampling distribution of the test statistics. The extension  of the testing procedures for spatially correlated data is presented in Section \ref{sec:gof_circular_dep}. Section \ref{sec:gof_circular_dep_cal} contains bootstrap approaches to calibrate the tests in this spatial framework. A simulation study for assessing the performance of the tests using PSCB and NPSCB methods is provided in Section \ref{sec:gof_circular_dep_simu}. Finally, {some conclusions and ideas for further research} are provided in Section \ref{sec:conclusions_gof_c}.

\section[Goodness-of-fit tests for linear-circular regression models]{{Goodness-of-fit tests} for circular regression models {with independent data}}\label{sec:gof_circular_ind}

Let $\{(\mathbf{X}_i,\Theta_i) \}_{i=1}^{n}$ be a random sample from  $(\mathbf{X},\Theta)$, where $\Theta$ is a circular random variable taking values on $\mathbb{T}=[0,2\pi)$, and $\mathbf{X}$ is a random variable with density $f$ {and support on} $\mathcal{D}\subseteq\mathbb{R}^d$. Assume that {the following regression model holds}:
\begin{equation}\label{C_model}
\Theta_i=[m(\mathbf{X}_i)+{\varepsilon}_i](\mbox{\texttt{mod}} \, 2\pi), \quad i=1,\dots,n,
\end{equation} 
where $m$ is a {circular regression} function, and {$\varepsilon_i, i=1,\ldots,n$, is an independent sample of a circular variable $\varepsilon$,} with zero mean direction and  finite concentration. This implies that ${\mathbb{E}}[\sin ({\varepsilon})\mid\mathbf{X}=\mathbf{x}]=0$. Additionally, the following notation is used: $\ell(\mathbf{x}) ={\mathbb{E}}[\cos ({\varepsilon})\mid\mathbf{X}=\mathbf{x}]$,  $\sigma^2_1(\mathbf{x})={\mathbb{V}{\rm ar}}[\sin({\varepsilon})\mid\mathbf{X}=\mathbf{x}]$,
$\sigma^2_2(\mathbf{x})={\mathbb{V}{\rm ar}}[\cos({\varepsilon})\mid\mathbf{X}=\mathbf{x}]$ and
$\sigma_{12}(\mathbf{x})={\mathbb{E}}[\sin({\varepsilon})\cos({\varepsilon})\mid\mathbf{X}=\mathbf{x}]$.

Considering the regression model (\ref{C_model}), one of the aims of the present {research} is to propose and study different testing procedures to assess the suitability of a general class of parametric circular regression models.
{Specifically, in this work, we focus on the following testing problem:
\begin{equation}\label{eq:test_circular}
H_0:m\in \mathcal{M}_{\bm{\beta}}=\{\beta_0+g(\bm{\beta}^T_1\mathbf{X}), \beta_0\in\mathbb{R}, \bm{\beta}_1,\in\mathbb{R}^d\} \hspace{0.8cm}\text{vs.} \hspace{0.8cm} H_a:m\notin \mathcal{M}_{\bm{\beta}},
\end{equation}
where $g$ is a link function {mapping} the real line onto the circle.}
  As pointed out in Section \ref{sec:gof_circular_int}, the procedure proposed in this work consists {in} comparing {a} ({non-smoothed} or smoothed) parametric {fit} {with a nonparametric} {estimator of the circular regression function $m$}, measuring the circular distance between both fits and employing this distance {as a} test statistic.  The parametric and nonparametric estimation methods considered in this proposal are described in {what follows}.  

\subsection{Parametric circular regression estimator}
\label{sec:gof_circular_estimation_par}

{As mentioned in the {Introduction}, our proposal requires a parametric estimator of the circular regression function {$m$}, once a parametric family is set as the null hypothesis.} Notice that, for instance, the procedures based on least squares for  Euclidean data,  are not appropriate when the response variable is of
circular nature. Minimizing the sum of squared differences between the observed and predicted
values may lead to erroneous results, since the squared
difference is not an appropriate measure on the circle. 

A circular {analog} to
least squares regression  for  models with  a circular response and a set of Euclidean covariates was presented by \citet{lund1999least}. Specifically, assume the regression model (\ref{C_model}) holds and consider ${m}\in \mathcal{M}^c_{\bm{\beta}}=\{{m}_{\bm{\beta}},{\bm{\beta}}\in\bm{\mathcal{B}}\}$, where ${m}_{\bm{\beta}}$ is a certain parametric circular regression model with parameter vector $\bm{\beta}$. A parameter estimate of $\bm{\beta}$ could be obtained by minimizing the sum of the circular distances between the observed and predicted values as follows: 
\begin{eqnarray}
\hat{{\bm{\beta}}}={\rm arg}\min_{\bm{{\bm{\beta}}}}\sum_{i=1}^n \left\{  1-\cos\left[\Theta_i-{{m}}_{\bm{\beta}}(\mathbf{X}_i)\right]\right\}.\label{beta_circular_est_LS}\end{eqnarray}

The value of the parameter minimizing the previous expression will be used to construct the parametric circular regression estimator, namely,  ${m}_{\hat{\mathbf{\beta}}}$.

An equivalent parameter estimator can be obtained using a maximum-likelihood approach {\citep{lund1999least}}. If it is assumed that the response variable (conditional on $\mathbf{X}$) has a von Mises distribution with mean direction given by ${{m}}_{\bm{\beta}}$ and concentration parameter $\kappa$, the maximum likelihood estimator of ${{m}}_{\bm{\beta}}$ maximizes the following expression
\begin{eqnarray}
\sum_{i=1}^n \cos\left[\Theta_i-{{m}}_{\bm{\beta}}(\mathbf{X}_i)\right].\label{beta_circular_est_ML}\end{eqnarray}

Notice that the circular least squares estimator given in (\ref{beta_circular_est_LS}) also maximizes the expression (\ref{beta_circular_est_ML}) and, therefore, as pointed out before, assuming a von Mises distribution,  the circular least squares estimator coincides with the maximum likelihood estimator. For further details see \citet{lund1999least}.

Assuming that the response variable follows a von Mises distribution and considering {as $\mathcal{M}^c_{\bm{\beta}}$ the parametric family $\mathcal{M}_{\bm{\beta}}$ given in (\ref{eq:test_circular})}, an iteratively reweighted least squares algorithm can be used to compute the maximum likelihood estimators of $\kappa$, $\beta_0$ and  $\bm{\beta}_1$ \citep[see][]{lund1999least,fisher1992regression}. The extension of these results to the case of a generic parametric family {has not been explicetly considered.}

\subsection{Nonparametric circular regression estimator}
\label{sec:gof_circular_estimation_npar}

{A nonparametric regression} estimator for {$m$ in} model (\ref{C_model}) is presented in this section. Notice that the circular regression function $m$ is the conditional mean direction of $\Theta$ given $\mathbf X$ which, at a point $\mathbf x$, can be defined as the minimizer of the  risk ${\mathbb{E}}\{1-\cos[\Theta-m(\mathbf{X})]\mid \mathbf X=\mathbf{x}\}$.  Specifically, the minimizer of this cosine risk  is given by $m(\mathbf{x})=\mbox{atan2}[m_1(\mathbf{x}),m_2(\mathbf{x})]$, where  $m_1(\mathbf{x})={\mathbb{E}}[\sin(\Theta)\mid\mathbf{X}=\mathbf{x}]$ and $m_2(\mathbf{x})={\mathbb{E}}[\cos(\Theta)\mid\mathbf{X}=\mathbf{x}]$. Therefore, replacing $m_1$ and $m_2$ by appropriate estimators, an estimator for $m$ can be directly obtained. In particular, a whole class of kernel-type estimators for $m$ at $\mathbf x\in \mathcal{D}$ can be defined by considering local polynomial estimators for $m_1(\mathbf x)$ and $m_2(\mathbf x)$. Specifically,  estimators of the form:
\begin{equation}\label{C_est}
{\hat{m}}_{\mathbf{H}}(\mathbf{x};p)=\mbox{atan2}[\hat{m}_{1, \mathbf{H}}(\mathbf{x};p),\hat{m}_{2, \mathbf{H}}(\mathbf{x};p)]
\end{equation}
are considered, where for any integer $p\geq 0$, $\hat m_{1, \mathbf{H}}(\mathbf{x};p)$ and $\hat m_{2, \mathbf{H}}(\mathbf{x};p)$ denote  the $p$th order local polynomial estimators (with bandwidth matrix $\mathbf H$) of $ m_1(\mathbf{x})$ and $m_2(\mathbf{x})$, respectively. The special cases $p=0$ and $p=1$ yield a Nadaraya--Watson (or local constant) type estimator and  a local linear type  estimator of $m(\mathbf x)$, respectively. Asymptotic properties of these estimators, considering model (\ref{C_model}), {have been studied by} \citet{meilan2019c}.

\subsection{{The test statistics}}
\label{sec:gof_circular_testing}
In this section, in order to check if the circular regression function belongs to a general class of parametric models, goodness-of-fit tests are presented. 
{We consider the testing problem (\ref{eq:test_circular}).}

Test statistics  to address (\ref{eq:test_circular}) are proposed and studied. The first approach considers a weighted circular distance between the nonparametric and parametric fits: 
\begin{equation}
\label{eq:statistic_circular_1}
T^{1}_{n,p}=\int_{\mathcal{D}}^{}\{1-\cos[{\hat{m}}_{\mathbf{H}}(\mathbf{x};p)-{{m}}_{\hat{\bm{\beta}}}(\mathbf{x})]\}w(\mathbf{x})d\mathbf{x},
\end{equation}
for $p=0,1$, where $w$ is a weight function that helps in mitigating possible boundary effects.
The estimators ${\hat{m}}_{\mathbf{H}}(\mathbf{x};p)$, for $p=0,1$, are the Nadaraya-Watson or the local linear type estimators of the circular regression function {$m$}, given in (\ref{C_est}). The parametric estimator ${{m}}_{\hat{{\bm{\beta}}}}$ was  described in Section  \ref{sec:gof_circular_estimation_par}. 

The second approach is similar to the first one, {but considering} a smooth version of the parametric fit:
\begin{equation}
\label{eq:statistic_circular_2}
T^{2}_{n,p}=\int_{\mathcal{D}}^{}\{1-\cos[{\hat{m}}_{\mathbf{H}}(\mathbf{x};p)-{\hat{m}}_{\mathbf{H},\hat{\bm{\beta}}}(\mathbf{x};p)]\}w(\mathbf{x})d\mathbf{x},
\end{equation}where ${\hat{m}}_{\mathbf{H},\hat{\bm{\beta}}}(\mathbf{x};p)$, for $p=0,1$, are smooth versions of the parametric estimator ${{m}}_{\hat{{\bm{\beta}}}}$, which are given by:
\begin{equation*}\label{C_est_smooth}
{\hat{m}}_{\mathbf{H},\hat{\bm{\beta}}}(\mathbf{x};p)=\mbox{atan2}[\hat{m}_{1, \mathbf{H},\hat{\bm{\beta}}}(\mathbf{x};p),\hat{m}_{2, \mathbf{H},\hat{\bm{\beta}}}(\mathbf{x};p)],
\end{equation*}
with
\begin{equation*}
\label{C_estNW_smooth}\hat m_{j, \mathbf{H},\hat{{\bm{\beta}}}}(\mathbf{x};0)=\left\{\begin{array}{lc}\dfrac{\sum_{i=1}^n K_{\mathbf{H}}(\mathbf{X}_i-\mathbf{x})\sin[{m}_{\hat{\bm{\beta}}}(\mathbf{X}_i)]}{\sum_{i=1}^n K_{\mathbf{H}}(\mathbf{X}_i-\mathbf{x})}&\text{if $j=1$},\\\\ \dfrac{\sum_{i=1}^n K_{\mathbf{H}}(\mathbf{X}_i-\mathbf{x})\cos[{m}_{\hat{\bm{\beta}}}(\mathbf{X}_i)]}{\sum_{i=1}^n K_{\mathbf{H}}(\mathbf{X}_i-\mathbf{x})}&\text{if $j=2$},\end{array}\right.
\end{equation*}
\begin{equation*}\label{C_estLL_smooth}
\hat m_{j, \mathbf{H},\hat{{\bm{\beta}}}}(\mathbf{x};1)=\left\{\begin{array}{lc}\mathbf{e}_1^T(\bm{\mathcal{X}}_{\mathbf{x}}^T\bm{\mathcal{W}}_{\mathbf{x}}\bm{\mathcal{X}}_{\mathbf{x}})^{-1}\bm{\mathcal{X}}_{\mathbf{x}}^T\bm{\mathcal{W}}_{\mathbf{x}}\hat{\mathbf{S}}&\text{if $j=1$},\\\\ \mathbf{e}_1^T(\bm{\mathcal{X}}_{\mathbf{x}}^T\bm{\mathcal{W}}_{\mathbf{x}}\bm{\mathcal{X}}_{\mathbf{x}})^{-1}\bm{\mathcal{X}}_{\mathbf{x}}^T\bm{\mathcal{W}}_{\mathbf{x}}\hat{\mathbf{C}}&\text{if $j=2$},\end{array}\right.
\end{equation*}
being $\hat{\mathbf{S}}=(\sin[{m}_{\hat{\bm{\beta}}}(\mathbf{X}_1)],\dots,\sin[{m}_{\hat{\bm{\beta}}}(\mathbf{X}_n)])^T$ and $\hat{\mathbf{C}}=(\cos[{m}_{\hat{\bm{\beta}}}(\mathbf{X}_1)],\dots,\cos[{m}_{\hat{\bm{\beta}}}(\mathbf{X}_n)])^T$. 

In order to formally {address problem (\ref{eq:test_circular})} using the test statistics $T_{n,p}^{1}$ and $T_{n,p}^{2}$ given in   (\ref{eq:statistic_circular_1}) and in (\ref{eq:statistic_circular_2}), respectively, it is essential to approximate the distribution of the
test statistic under the null hypothesis. {Deriving the asymptotic distribution of the statistics is out of the scope of this work. However, some guidelines to compute these expressions are provided in Section \ref{sec:conclusions_gof_c}. For the application in practice of our proposal, the} distribution of the tests under the null hypothesis is approximated using bootstrap procedures and  analyzed through an empirical study.  

\begin{figure}[t]
	\centering
	\includegraphics[width=1\textwidth]{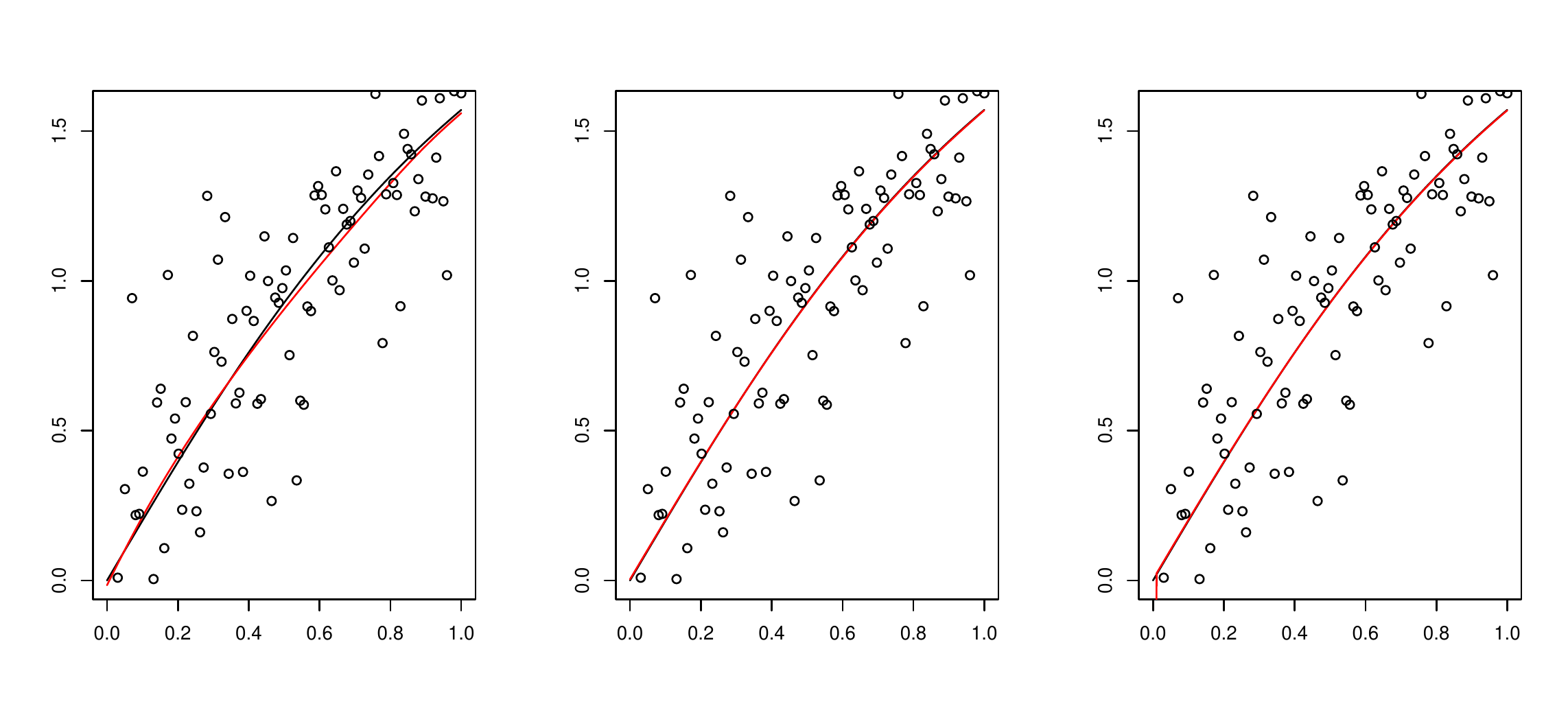}
	\caption{ Red lines:   local linear type regression estimator (left),  parametric fit (center) and smooth version of the parametric fit (right), with sample points and  circular regression function (black lines). Sample of size $n=100$ generated on the unit interval, following model (\ref{C_model}), with regression function  (\ref{trend_circ_sim_uni}), for $c=0$, and circular errors $\varepsilon_i$ drawn from a $vM(0,10)$.}
	\label{fits_circular}
\end{figure}

If the null hypothesis in the testing  problem given in (\ref{eq:test_circular}) holds, then the ({non-smoothed} or smoothed) parametric fit and the nonparametric circular regression estimator   will be {similar} and, therefore, the value of the test statistics   $T_{n,p}^{1}$ and $T_{n,p}^{2}$ will be {relatively} {small}. Conversely, if the null hypothesis {does not hold}, the fits will be {different} and the value of $T_{n,p}^{1}$ and $T_{n,p}^{2}$  will be {fairly} large. So, the null hypothesis will be  rejected if the circular distance between {both} fits exceeds a critical value.  {For a visual} illustration of the performance of the tests (where, for the sake of simplicity, a {model with a} single covariate is considered), suppose that a sample of size $n=100$ is generated following model (\ref{C_model}), with regression function (\ref{trend_circ_sim_uni}), with $c=0$, and  random errors  $\varepsilon_i$ drawn from a von Mises distribution $vM(0,10)$. If we want to test if ${m}({X})\in\{\beta_0+2\rm{atan}(\beta_1X), \beta_0,\beta_1\in \mathbb{R}\},$
using the test statistics given in (\ref{eq:statistic_circular_1}) and in (\ref{eq:statistic_circular_2}),  the estimators  ${\hat{m}}_{h}({x};p)$, ${{m}}_{\hat{\bm{\beta}}}({x})$ and ${\hat{m}}_{h,\hat{\bm{\beta}}}({x};p)$ (denoting by $h$ the smoothing parameter when $d=1$)   must be computed. In this case, the estimator obtained from (\ref{beta_circular_est_LS}) is considered for the parametric fit. The local linear type estimator ${\hat m}_{h}({x};1)$ given in (\ref{C_est}) is
employed to compute the nonparametric counterpart. A  triweight kernel and the
optimal bandwidth obtained by minimizing the circular average squared error (CASE), defined as:
\begin{equation}\label{CASE}
{\rm CASE}[{\hat{m}}_{\mathbf{H}}(\mathbf{x};p)]=\frac{1}{n}\sum_{i=1}^n \left\{1- \cos\left[m(\mathbf{X}_i) - {\hat{m}}_{\mathbf{H}}(\mathbf{X}_i;p)\right]\right\},\end{equation}
for $p=0,1$ and $d=1$, are considered to compute ${\hat{m}}_{h}({x};1)$ and ${\hat{m}}_{h,\hat{\bm{\beta}}}({x};1)$. Figure \ref{fits_circular} shows  in red lines the  local linear type regression estimator (left panel),  the parametric fit (center panel) and the smooth version of the parametric fit (right panel), with sample points and the  circular regression function (black lines).  It seems that all estimates  are very {similar} and, therefore, the value of the test statistics $T_{n,p}^{1}$ and $T_{n,p}^{2}$ should be {small},  and consequently, there would possibly be no evidences against the assumption that the circular regression function {belonging} to the parametric family ${{m}}_{\bm{\beta}}({X})=\beta_0+2\rm{atan}(\beta_1X)$.

Notice that the test statistics given in (\ref{eq:statistic_circular_1}) and in (\ref{eq:statistic_circular_2}), respectively,  {depend on} the bandwidth matrix $\mathbf{H}$ (or on the bandwidth parameter $h$, if $d=1$). A non-trivial problem in goodness-of-fit testing is the bandwidth choice, since
the optimal bandwidth for estimation may not be the optimal one for testing (being
not even clear what optimal means). For instance, \citet{fan2001generalized}, \citet{eubank2005testing} and \citet{hart2013nonparametric}  gave some strategies on bandwidth selection in testing problems. This issue  was also discussed further in detail by \citet{sperlich2013comments}. {As usual in the context of goodness-of-fit tests for regression based on nonparametric smoothers, the performance of the test statistics will be analyzed for a range of bandwidths, in order to evaluate the impact of this parameter in the numerical results.}

\subsection{Calibration in practice}\label{sec:gof_circular_ind_cal}
Once a suitable test statistic is available, {in order to solve the testing problem {(\ref{eq:test_circular})},  a procedure for calibration of critical values is required}. This task can be done
by means of bootstrap resampling {algorithms}.

In what follows, a  description of two different bootstrap proposals designed {to approximate the distribution  (under the null hypothesis) of the tests statistics 
given in} (\ref{eq:statistic_circular_1}) and in (\ref{eq:statistic_circular_2}) for independent data (PCB and NPCB) are
presented. The main difference between them is the mechanism employed to obtain the residuals. {As noted in the Introduction, the residuals in PCB} come from the parametric regression  estimator. On the other hand, {for the NPCB algorithm}, the residuals
are obtained from the nonparametric regression estimator. In order to present the PCB and NPCB  resampling methods, a generic bootstrap algorithm is described. No matter the method used,  ${\hat{m}}$  denotes the parametric or the nonparametric circular regression estimator.

\begin{algorithm}[H]{}
	\caption{}
	\begin{algorithmic}
		\State 1. Compute {the parametric or the nonparametric regression estimates (described in Sections  \ref{sec:gof_circular_estimation_par} and \ref{sec:gof_circular_estimation_npar}}, respectively), namely ${\hat{m}}(\mathbf X_i)$, $i=1,\ldots,n$, depending if a parametric (PCB) or a  nonparametric (NPCB) bootstrap procedure is employed.  
		\State 2.	From the residuals    $\hat{\varepsilon}_i=[\Theta_i-{\hat{m}}(\mathbf{X}_i)](\mbox{\texttt{mod}} \, 2\pi)$, $i=1,\ldots,n$, draw independent bootstrap  residuals, $\hat{\varepsilon}^*_i$, $i=1,\ldots,n$.
		\State 3. Bootstrap samples are $\{(\mathbf{X}_i,\Theta^*_i)\}_{i=1}^n$ with  $\Theta^*_i=[{m}_{\hat{\mathbf{\beta}}}(\mathbf{X}_i)+\hat{\varepsilon}^*_i](\mbox{\texttt{mod}} \, 2\pi)$, being ${m}_{\hat{\mathbf{\beta}}}(\mathbf X_i)$  the parametric regression estimator under $H_0$. 
		\State 4.  Using the bootstrap sample $\{(\mathbf{X}_i,\Theta^*_i)\}_{i=1}^n$, the bootstrap test statistics $T^{1,*}_{n,p}$ and $T^{2,*}_{n,p}$ are computed as in (\ref{eq:statistic_circular_1}) and in (\ref{eq:statistic_circular_2}).
		\State 5. Repeat Steps 2-4 a large number of times $B$.
	\end{algorithmic}
	\label{alg_2}
\end{algorithm}

 In Step 1 of the previous algorithm, {in the PCB approach,} the circular regression function is estimated parametrically, employing  the procedure described in Section \ref{sec:gof_circular_estimation_par}. {Alternatively},  the NPCB  tries to avoid possible misspecification problems by using more flexible regression estimation methods than those employed in PCB. {Then, using the same arguments as in \citet{gonzalez1993testing} to increase the power of the test, in the NPCB method, the nonparametric circular regression estimator given in (\ref{C_est}) is employed in} Step 1 of the bootstrap Algorithm \ref{alg_2}. 

Notice that the empirical distribution of the $B$ bootstrap test statistics can be employed to approximate  the finite sample distribution of the  test statistics $T^{1}_{n,p}$ and $T^{2}_{n,p}$, under the null hypothesis. {Denoting} by {$\{T_{n,p,1}^{j,*},{\ldots},T_{n,p,B}^{j,*}\}$ (for $j=1,2$) the sample of the  $B$ bootstrap test statistics given in (\ref{eq:statistic_circular_1}) {and in (\ref{eq:statistic_circular_2})},  and defining its $(1-\alpha)$ quantile $t_{\alpha,p}^{j,*}$,  the null hypothesis in (\ref{eq:test_circular}) will be rejected if $T^{j}_{n,p}>t_{\alpha,p}^{j,*}$. Additionally, the $p$-{values} of the test {statistics} {($j=1,2$)} can be approximated by:
\begin{equation}\label{pvalue}p\mbox{-value}=\dfrac{1}{B}\sum_{b=1}^B\mathbb{I}_{\{T_{n,{p,}b}^{j,*}>T^{j}_{n,p}\}}.\end{equation}}

\subsection{Simulation study}\label{sec:gof_circular_ind_simu}
The finite sample performance of the proposed tests, using the bootstrap approaches described in Algortihm \ref{alg_2} for their calibration, is illustrated in this section with a simulation study, {considering a regression model} {with a single real-valued {covariate} and {also} with a bidimensional one}.

\subsubsection{{Simulation} experiment with a single covariate}
In order to study the performance of the proposed tests considering a regression model with  a circular response and a {single real-valued} covariate, the {parametric} regression family
$
{\mathcal{M}}_{1,\bm{\beta}}=\{\beta_0+2\rm{atan}(\beta_1X),\beta_0, \beta_1\in\mathbb{R}\}$ {is chosen}. For different values of $c$ the regression function
\begin{equation}
{m}(X)=2\rm{atan}(X)+c\cdot\rm{asin}(2X^5-1)
\label{trend_circ_sim_uni}
\end{equation}
is considered. Therefore, the parameter $c$ controls whether the null ($c=0$) or the alternative ($c\neq 0$) hypotheses {hold in problem (\ref{eq:test_circular})}. Values $c=0$, $1$, and $2$  are considered in the study. For each value of $c$, 500 samples of sizes  $n=50, 100$ and $200$ are generated on the unit interval, following model (\ref{C_model}){ with regression function (\ref{trend_circ_sim_uni})} and  circular errors $\varepsilon_i$ drawn {independently} from a von Mises distribution $vM(0,\kappa)$,  for different values of $\kappa$ (5, 10 and 15).

To analyze the behavior of the test statistics given in (\ref{eq:statistic_circular_1}) and in (\ref{eq:statistic_circular_2}) in the different scenarios, the bootstrap procedures described in Section \ref{sec:gof_circular_ind_cal} are applied, using  $B=500$ replications.   The {non-smoothed or smoothed} parametric {fits} used for constructing (\ref{eq:statistic_circular_1}) and (\ref{eq:statistic_circular_2}) {are} computed using the {procedures} given in {Sections \ref{sec:gof_circular_estimation_par} and \ref{sec:gof_circular_testing}, respectively}. The nonparametric fit is obtained using the estimator given in (\ref{C_est}), for $p=0,1,$ with a triweight kernel.  We  address the bandwidth selection
problem by using the same  procedure as the one used in \citet{hardle1993comparing}, \citet{alcala1999goodness}, {or} \citet{opsomer2010finding}, among others, applying the tests on a grid of several bandwidths. In order to use a reasonable grid of bandwidths, the optimal bandwidth selected by minimizing the CASE given in (\ref{CASE}), for $d=1$,  is calculated for each scenario. In this case, the values of the CASE optimal bandwidths are in the interval $[0.2, 0.6]$. Therefore,  the values of the bandwidth parameter $h=0.15, 0.25, 0.35, 0.45, 0.55, 0.65,$ are considered to compute both test statistics (\ref{eq:statistic_circular_1}) and (\ref{eq:statistic_circular_2}). The weight function used in both tests is $w(x)=\mathbb{I}_{\{x\in[1/\sqrt{n},1-1/\sqrt{n}]\}}$, to avoid possible boundary effects.

{\emph{Effect of sample size.}} Proportions of rejections of the null hypothesis, for a significance level $\alpha= 0.05$, considering $\kappa=10$, and different sample sizes, are shown in  Tables~\ref{table::c_simu_n_m1_ss_uni} and \ref{table::c_simu_n_m1_uni}, when using $T^{1}_{n,p}$ and $T^{2}_{n,p}$, respectively.    If $c=0$ (null hypothesis) and using the Nadaraya--Watson  {type} estimator,  the {proportions of rejections are} certainly much {lower} than the expected {values}.  Using this estimator, the test works fairly well when PCB is employed. NPCB provides really bad results.  When the local linear {type} estimator is used, the proportions of rejections are similar to the theoretical level, although these proportions are quite affected by the value of $h$.   For alternative assumptions ($c=1$ and $c=2$), {as expected,} as the sample size increases the {proportions} {of rejections} {are} larger {and {increase} with $c$}.  
 As pointed out before, substantial differences have been found when the local linear type estimator is employed, providing more satisfactory results than those obtained when the Nadaraya--Watson fit is used. Using the local linear estimator, NPCB presents a slightly better performance than PCB.   Although both test statistics provide a similar  behavior of the testing procedure, $T^{2}_{n,p}$ seems to give slightly better results. 

{\emph{Effect of $\kappa$.}} The performance of the tests {(for $\alpha=0.05$)} is studied for $n=200$ and for different values of the concentration parameter $\kappa$ in Tables \ref{table::c_simu_k_m1_ss_uni} and \ref{table::c_simu_k_m1_uni}, when using $T^{1}_{n,p}$ and $T^{2}_{n,p}$, respectively.  If $c=0$ and considering the local linear {type} estimator, the proportions {of rejections} are similar to the theoretical level when using both bootstrap approaches (PCB and NPCB). Results obtained when the Nadaraya--Watson fit is used are {quite poor}, specially when NPCB is employed.  For alternative assumptions, 
as expected, large values of the concentration parameter $\kappa$ lead to an increase in power, which justifies the correct performance of the bootstrap procedures. Considerable differences have been found if the Nadaraya--Watson or local linear  type estimators are employed in the test statistics, {especially} when NPCB is used.

\subsubsection{{Simulation} experiment with several covariates}
\label{sec:seve}
The extension for regression models with a circular {response} and two covariates is analyzed in this section. For this purpose, the {parametric regression family}
$
{\mathcal{M}}_{2,\bm{\beta}}=\{\beta_0+2\rm{atan}(\beta_1X_1+\beta_2X_2),\beta_0, \beta_1,\beta_2\in\mathbb{R}\}$ is chosen, 
and for different values of $c$ the regression function
\begin{equation}
{m}(\mathbf X)=2\rm{atan}(-X_1+X_2)+c\cdot\rm{asin}(2X_1^3-1),
\label{trend_circ_sim}
\end{equation}
{being $\textbf{X}=(X_1,X_2)$}, is considered.   For each value of $c$ ($c=0$, $1$, and $2$), 500 samples of sizes  $n=100, 225$ and $400$ are generated on a bidimensional regular grid in the unit square, following model (\ref{C_model}), with regression function (\ref{trend_circ_sim}) and  circular errors $\varepsilon_i$ drawn from a von Mises distribution $vM(0,\kappa)$, for  $\kappa=5, 10$ and $15$. The bootstrap procedures described in Section \ref{sec:gof_circular_ind_cal} are applied, using  $B=500$ replications.   The {non-smoothed or smoothed} parametric {fits} used for constructing (\ref{eq:statistic_circular_1}) and (\ref{eq:statistic_circular_2}) {are} computed using the {procedures} given in {Sections \ref{sec:gof_circular_estimation_par} and \ref{sec:gof_circular_testing}, respectively}. The nonparametric fit is obtained using the estimator given in (\ref{C_est}), for $p=0,1,$ with a multiplicative triweight kernel.   In order to simplify the calculations, the bandwidth matrix is restricted to a class of  diagonal matrices with both equal elements. {In this case, t}he diagonal elements of the CASE optimal bandwidths are in the interval $[0.3, 0.8]$. Therefore,  diagonal {bandwidth} matrices $\mathbf{H}=\text{diag}(h,h)$ with different  values of $h$, $h= 0.25, 0.35, 0.45, 0.55, 0.65, 0.75, 0.85,$ are considered to compute both test statistics (\ref{eq:statistic_circular_1}) and (\ref{eq:statistic_circular_2}). The weight function used in both tests is $w(\mathbf x)=\mathbb{I}_{\{\mathbf x\in[1/\sqrt{n},1-1/\sqrt{n}]\times[1/\sqrt{n},1-1/\sqrt{n}]\}}$, to avoid possible boundary effects.

{\emph{Effect of sample size.}} Proportions of rejections of the null hypothesis, for a significance level $\alpha= 0.05$, considering $\kappa=10$, and different sample sizes, are shown Tables~\ref{table::c_simu_n_m1_ss}, when using $T^{1}_{n,p}$.   It can be observed that using both {bootstrap} methods {(PCB and NPCB)}, the test has a reasonable behavior. If $c=0$ (null hypothesis) and considering the local linear estimator, the {proportions} of rejections are similar to the theoretical level. As for a single covariate, results using the Nadaraya--Watson {type} estimator and NPCB are really bad.   For alternative assumptions ($c=1$ and $c=2$), NPCB presents a slightly better performance than the PCB, when using the local linear {type} estimator. Notice that,  in most of the cases, an increasing power of the test when the values of $h$ increase is observed.  For all the scenarios, the power of the test becomes larger as the value of $c$ increases. Again, considerable differences have been found when the local linear type estimator is employed. {Similar} conclusions {to} those given for  $T^{1}_{n,p}$ were obtained when the test statistic  $T^{2}_{n,p}$ was employed {(see Table~\ref{table::c_simu_n_m1})}.  

{\emph{Effect of $\kappa$.}} The performance of the bootstrap procedures is analyzed for $n=400$ and for different values of the concentration parameter $\kappa$ when using $T^{1}_{n,p}$, {for $\alpha=0.05$, in  Table \ref{table::c_simu_k_m1_ss}}. If $c=0$, the proportions {of rejections} are similar to the theoretical level when using both bootstrap approaches (PCB and NPCB). For larger values of the concentration parameter $\kappa$, the bandwidth values providing an effective calibration must be smaller. For alternative assumptions, 
if the value of the concentration parameter $\kappa$ is larger, an increasing power is obtained. In almost all scenarios,  some differences have been found if the Nadaraya--Watson or {the} local linear  type estimators are employed in the test statistics. Results considering the test statistic $T^{2}_{n,p}$  are summarized in Table \ref{table::c_simu_k_m1}. Similar conclusions to those provided for $T^{1}_{n,p}$  were obtained.

\section[Goodness-of-fit tests for spatial-circular regression models]{{Goodness-of-fit tests} for circular regression models with spatially correlated data}\label{sec:gof_circular_dep}
The testing problem (\ref{eq:test_circular}) is addressed in Section \ref{sec:gof_circular_ind} for independent data, by constructing  weighted circular test statistics.  In this section, these test statistics are also analyzed considering a linear-circular regression model with spatially correlated errors. 

Assume the linear-circular regression
model given in (\ref{C_model}), but supposing that the circular errors are spatially correlated. More specifically, we consider the  linear-circular regression model given in (\ref{C_model}):
\begin{equation}\label{model_spatial}
\Theta_i=[m(\mathbf{X}_i)+{\varepsilon}_i](\mbox{\texttt{mod}} \, 2\pi), \quad i=1,\dots,n,
\end{equation} 
where $m$ is a smooth trend or regression function and the $\varepsilon_i$ are random angles,  such that, ${\mathbb{E}}[\sin (\varepsilon_i)\mid\mathbf{X}=\mathbf{x}]=0$, and additionally, satisfying in this dependence framework that
\begin{eqnarray*}
	{\mathbb{C}\rm ov}[\sin(\varepsilon_i),\sin(\varepsilon_j)\mid\mathbf{X}_i, \mathbf{X}_j]&=&\sigma^2_1\rho_{1,n}(\mathbf{X}_i-\mathbf{X}_j),\\
	{\mathbb{C}\rm ov}[\cos(\varepsilon_i),\cos(\varepsilon_j)\mid\mathbf{X}_i, \mathbf{X}_j]&=&\sigma^2_2\rho_{2,n}(\mathbf{X}_i-\mathbf{X}_j),\\
	{\mathbb{C}\rm ov}[\sin(\varepsilon_i),\cos(\varepsilon_j)\mid\mathbf{X}_i, \mathbf{X}_j]&=&\sigma_{12}\rho_{3,n}(\mathbf{X}_i-\mathbf{X}_j),
\end{eqnarray*}
with    $\sigma^2_k<\infty$ for $k=1,2$, and $\sigma_{12}<\infty$. The continuous stationary correlation functions $\rho_{k,n}$ satisfy $\rho_{k,n}(\bm{0})=1$, $\rho_{k,n}(\mathbf{x})=\rho_{k,n}(-\mathbf{x})$, and $\abs{\rho_{k,n}(\mathbf{x})}\le1$, for any  $\mathbf{x}\in\mathcal{D}\subset\mathbb{R}^d$, and $k=1,2,3$.  {The subscript $n$ in $\rho_{k,n}$ indicates that the correlation functions vary with $n$ (specifically, the correlation functions are assumed to be short-range and shrink as $n$ goes to infinity). Note also that the subscript $k$ does not correspond to an integer sequence and it just indicates if the correlation corresponds to the sine process ($k=1$), the cosine process ($k=2$) or if it is the cross-correlation between them ($k=3$).
}

In order to solve the  testing problem (\ref{eq:test_circular}) {in this context}, the estimator described in Section \ref{sec:gof_circular_estimation_par} is {likewise} employed for the parametric fit. Probably, more accurate results would be obtained if an estimator taking the spatial dependence structure into account is considered. However, the problem of estimating parametrically the regression function accounting {for} the dependence structure (up to {our} knowledge)
has not been considered in the statistical literature. Some guidelines about a possible iterative least squares estimator ({taking} the possible spatial dependence structure {into account})  are provided in Section \ref{sec:conclusions_gof_c}. {Kernel-type} estimators given in (\ref{C_est}) are employed for the nonparametric fit. 
{These nonparametric estimators have been studied in \cite{meilan2020a} in the context of spatially correlated data.}

For illustration purposes, a sample of size  $n=400$ is generated on a bidimensional regular grid in the unit square, assuming the linear-circular regression model (\ref{model_spatial}), with regression function (\ref{trend_circ_sim}), being $c=0$. The circular spatially correlated errors{ $\varepsilon_i$, $i=1, \ldots,n$,} are drawn  from {a wrapped Gaussian spatial process} \citep{jona2012spatial}{ with the following steps}:
\begin{eqnarray*}\label{wrapped_gen}
	\varepsilon_i=Y_i(\texttt{mod}\, 2\pi), \quad i=1,\dots,n,
\end{eqnarray*}
where $\{Y_i=Y(\mathbf{X}_i),\;i=1,\ldots,n\}$ is a realization  of a real-valued Gaussian spatial process, where each observation can be decomposed as:
\begin{equation}\label{processdecomposedW}
Y_i=\mu+w_i, \quad i=1,\dots,n,
\end{equation} 
being $\mu=\mu(\mathbf{X}_i)$ the mean and $w_i$  random variables of a zero mean Gaussian spatial process with ${\mathbb{C}\rm ov}(w_i,w_j\mid\mathbf{X}_i, \mathbf{X}_j)=\sigma^2\rho_{n}(\mathbf{X}_i-\mathbf{X}_j)$. The variance of  $w_i$ is denoted by $\sigma^2$ and $\rho_n$ is a continuous stationary correlation function satisfying $\rho_n(\bm{0})=1$, $\rho_n(\mathbf{x})=\rho_n(-\mathbf{x})$, and $\abs{\rho_n(\mathbf{x})}\le1$, $\forall \mathbf{x}$. Note that {a} realization of {this} wrapped Gaussian {spatial} process can be {written} in vector form as $\bm{\varepsilon}=(\varepsilon_1,\ldots,\varepsilon_n)^T$, with mean direction vector $\mu \mathbf{1}_n$, being $\mathbf{1}_n$ a $n\times  1$ vector with every entry equal to 1, and covariance matrix  $\sigma^2 \mathbf{R}_n$, where $\mathbf{R}_n(i,j)=\rho_n(\mathbf{X}_i-\mathbf{X}_j)$ is the $(i,j)$-entry of the correlation matrix $\mathbf{R}_n$. 

\begin{figure}[t]
	\centering
	\includegraphics[width=0.75\textwidth]{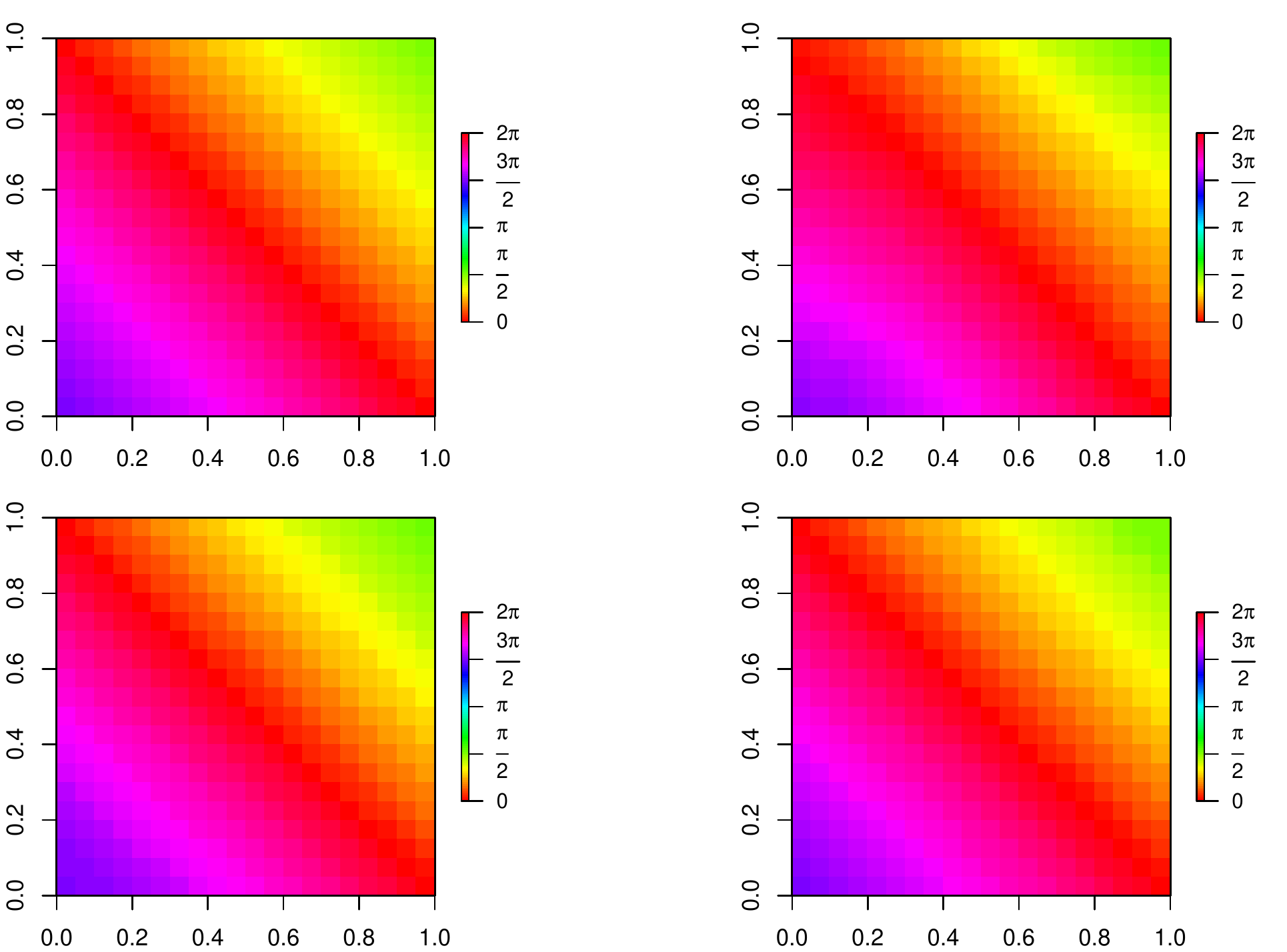}
	\caption{Circular regression function (top left),  the local linear type regression estimator (top right), the parametric fit (bottom left) and the  smooth version of the parametric fit (bottom right). Sample of size $n=400$ generated on a bidimensional regular grid in the unit square, following model (\ref{C_model}), with regression function  (\ref{trend_circ_sim}), for $c=0$, and circular errors $\varepsilon_i$ drawn from from wrapped Gaussian {spatial} processes with zero mean  and exponential covariance structure, given in (\ref{expo_w}), with  $\sigma^2=1$  and  $a_e=0.3$.}
	\label{fits_circular_spatial}
\end{figure}

In this {particular example}, the circular spatially correlated errors are drawn  from {a} wrapped Gaussian {spatial process, considering that, in (\ref{processdecomposedW}), $\mu=0$ and $w_i$ is a} zero mean {process with} exponential covariance structure:
\begin{equation}
{\mathbb{C}\rm ov}({w}_i,w_j \mid\mathbf{X}_i, \mathbf{X}_j)=   
\sigma^2[\exp(-\norm{\mathbf{X}_i-\mathbf{X}_j}/a_e)],
\label{expo_w}
\end{equation} with  $\sigma^2=1$  and  $a_e=0.3$. In order to test if ${m}(\mathbf{X})\in\{\beta_0+2\rm{atan}(\beta_1X_1+\beta_2X_2), \beta_0,\beta_1,\beta_2\in \mathbb{R}\},$
being $\textbf{X}=(X_1,X_2)$, using the test statistics given in (\ref{eq:statistic_circular_1}) and in (\ref{eq:statistic_circular_2}), ${\hat{m}}_{\mathbf{H}}(\mathbf{x};p)$, ${m}_{\hat{\bm{\beta}}}(\mathbf{x})$ and ${\hat{m}}_{\mathbf{H},\hat{\bm{\beta}}}(\mathbf{x};p)$  fits must be computed.  For the parametric counterpart, the estimator obtained from (\ref{beta_circular_est_LS}) is employed, while   local linear type estimators are
used for the nonparametric smoothers. A multiplicative triweight kernel and the
optimal bandwidth obtained by minimizing the CASE, given in (\ref{CASE}),
of the local linear type estimator are considered to compute ${\hat{m}}_{\mathbf{H}}(\mathbf{x};1)$ and ${\hat{m}}_{\mathbf{H},\hat{\bm{\beta}}}(\mathbf{x};1)$.  Figure \ref{fits_circular_spatial} shows  the theoretical circular regression function {(\ref{trend_circ_sim}), with $c=0$} (top left panel), the  local linear type regression estimator (top right panel), the parametric fit (bottom left panel) and the smooth version of the parametric fit (bottom right panel).  It can be seen that estimates at top right, bottom left and bottom right panels seem to be very {similar} and, therefore, the value of the test statistics $T_{n,p}^{1}$ and $T_{n,p}^{2}$ should be {small}. Consequently, the formal application of the tests will probably lead to assert that there is no evidences against the assumption that the regression function belongs to the parametric family ${{m}}_{\bm{\beta}}(\mathbf{X})=\beta_0+2\rm{atan}(\beta_1X_1+\beta_2X_2)$, with $\textbf{X}=(X_1,X_2)$. 

Practical methods to calibrate the test statistics $T_{n,p}^{1}$ and $T_{n,p}^{2}$ given in  (\ref{eq:statistic_circular_1}) and in (\ref{eq:statistic_circular_2}) for spatially correlated data are presented in the following section. 

\subsection{Calibration in practice}\label{sec:gof_circular_dep_cal}
This section is devoted to present bootstrap resampling methods  to calibrate in practice the test statistics $T_{n,p}^{1}$ and $T_{n,p}^{2}$ given in  (\ref{eq:statistic_circular_1}) and in (\ref{eq:statistic_circular_2}), respectively, considering the linear-circular regression model (\ref{model_spatial}) with spatially correlated errors.

The bootstrap Algorithm \ref{alg_2}, which was designed for independent data,  should not be used for spatial processes, as it does not account for the correlation
structure. The aim of this section  is to describe two different
proposals for test calibration which take the dependence of the data into account (PSCB and NPSCB).   The main difference between the proposals is how the resampling residuals
(required for mimicking the dependence structure of the errors) are computed. In PSCB (similarly to PCB), the residuals
are obtained from the parametric regression  estimator, while in NPSCB (analogously to NPCB), the residuals
are obtained from the nonparametric regression estimator. In both approaches,  in order to imitate the dependence structure of the errors, an appropriate spatial circular process model is  fitted to the residuals.

Next, a generic bootstrap algorithm is introduced to present the PSCB and NPSCB resampling approaches.  As in Algorithm \ref{alg_2}, no matter the method used, either
parametric or nonparametric, ${\hat{m}}$  denotes the parametric or the nonparametric circular regression estimator.

\begin{algorithm}[H]{}
	\caption{}
	\begin{algorithmic}
		\State 1.  Compute {the parametric or the nonparametric regression estimates (described in Sections  \ref{sec:gof_circular_estimation_par} and \ref{sec:gof_circular_estimation_npar}}, respectively), namely ${\hat{m}}(\mathbf X_i)$, $i=1,\ldots,n$, depending if a parametric (PSCB) or a  nonparametric (NPSCB) bootstrap procedure is employed.  
		\State 2.	From the residuals    $\hat{\varepsilon}_i=[\Theta_i-{\hat{m}}(\mathbf{X}_i)](\mbox{\texttt{mod}} \, 2\pi)$, $i=1,\ldots,n$, fit a  spatial circular process.
		\State 3. Generate a random sample from the fitted model, $\hat{\varepsilon}^*_i$, $i=1,\ldots,n$.
		\State 4. Bootstrap samples are $\{(\mathbf{X}_i,\Theta^*_i)\}_{i=1}^n$ with  $\Theta^*_i=[{m}_{\hat{\mathbf{\beta}}}(\mathbf{X}_i)+\hat{\varepsilon}^*_i](\mbox{\texttt{mod}} \, 2\pi)$, being ${m}_{\hat{\mathbf{\beta}}}(\mathbf X_i)$  the parametric regression estimator. 
		\State 5.  Using the bootstrap sample $\{(\mathbf{X}_i,\Theta^*_i)\}_{i=1}^n$, the bootstrap test statistics $T^{1,*}_{n,p}$ and $T^{2,*}_{n,p}$ are computed as in (\ref{eq:statistic_circular_1}) and in (\ref{eq:statistic_circular_2}).
		
		\State 6. Repeat Steps 3-5 a large number of times $B$.
	\end{algorithmic}
	\label{alg_3}
\end{algorithm}

Notice that Algorithm \ref{alg_3} is a modification of Algorithm \ref{alg_2}. Two additional steps are included in Algorithm \ref{alg_3} (Steps 2 and 3) trying to mimic properly the spatial dependence structure of the cirular errors   in the bootstrap procedure.

As pointed out in Section \ref{sec:gof_circular_ind_cal} for independent data, considering the test {statistics $T^{j}_{n,p}$ ($j=1,2$)}, given in (\ref{eq:statistic_circular_1}) {and in (\ref{eq:statistic_circular_2})}, the null hypothesis in (\ref{eq:test_circular}) will be rejected if $T^{j}_{n,p}>t_{\alpha,p}^{j,*}$, where $t_{\alpha,p}^{j,*}$ is the $(1-\alpha)$ quantile of  the sample of the  $B$ bootstrap test statistics $\{T_{n,p,1}^{j,*},{\ldots},T_{n,p,B}^{j,*}\}$.   Moreover,  the $p$-{values} of the test {statistics} can be approximated  as in (\ref{pvalue}).

\subsection{Simulation experiment}\label{sec:gof_circular_dep_simu}

The performance of the proposed  test statistics and the bootstrap procedures, described in Algorithm \ref{alg_3},  are analyzed in a simulation study. The parametric circular regression family ${\mathcal{M}}_{2,\bm{\beta}}$ given in Section {\ref{sec:seve}} is chosen, and for different values of $c$ ($c=0,1,2$), the regression function (\ref{trend_circ_sim}) is considered.

In this study, 500 samples of sizes  $n=100, 225$ and $400$ are generated on a bidimensional regular grid in the unit square, assuming the linear-circular regression model (\ref{C_model}), with regression function (\ref{trend_circ_sim}), but considering circular spatially correlated errors generated  from wrapped Gaussian {spatial} processes \citep{jona2012spatial}. The realizations of the circular (error) $\{\varepsilon_i,\;i=1,\ldots,n\}$ are generated considering {a}  zero mean {process} with {the} exponential covariance structure given in (\ref{expo_w}). The value of the variance $\sigma^2$  is fixed equal to one, and  different values of the range parameter are considered: $a_e=0.1, 0.3, 0.6.$

The performance of Algorithm \ref{alg_3} is analyzed in this section. Notice that Algorithm \ref{alg_2}, which was designed for independent observations, should not be used in a spatial framework. In order to illustrate this issue,  Tables \ref{table::c_simu_n_sp_ind_ss} and \ref{table::c_simu_n_sp_ind} show the {proportions} of rejections of the null hypothesis for different sample sizes and $\alpha=0.05$, when using $T_{n,p}^{1}$ and $T_{n,p}^{2}$, respectively, for $p=0,1${ when using Algorithm \ref{alg_2}, but the circular errors are spatially correlated as explained before}. Considering both test statistics, it may seem that PCB and NPCB  present a good behavior in terms of power. However, the proportions of rejections under the null hypothesis are very large.  Results for $n=400$ and different spatial dependence degrees {(controlled by the range parameter, $a_e$)} are summarized in Tables \ref{table::c_simu_ae_sp_ind_ss} and \ref{table::c_simu_ae_sp_ind}, when using $T_{n,p}^{1}$ and $T_{n,p}^{2}$, respectively, for $p=0,1$. Again, it can be obtained that the {tests} do not  work properly under the null hypothesis.

The bootstrap procedures described in Algorithm \ref{alg_3} are now applied, using  $B=500$ replications. As pointed out previously, the test statistics $T_{n,p}^{1}$ and $T_{n,p}^{2}$ given in (\ref{eq:statistic_circular_1}) and in (\ref{eq:statistic_circular_2}), are computed using
the { non-smoothed or smoothed} parametric {fits} given in {Sections \ref{sec:gof_circular_estimation_par} and  \ref{sec:gof_circular_testing}, respectively}, while  the nonparametric fit is obtained using the estimator given in (\ref{C_est}), for $p=0,1,$ with a multiplicative triweight kernel. In practice, in order to implement  the bootstrap Algorithm \ref{alg_3}, a wrapped Gaussian spatial  process model is employed in Step 2. Following the proposal by \citet{jona2012spatial}, the model fitting within a Bayesian framework is performed using a Markov Chain Monte Carlo method. Assuming a linear Gaussian spatial process of the form (\ref{processdecomposedW}), to perform a Bayesian fit of the model, priors are needed for the model parameters. The authors suggest a normal prior for $\mu$, a truncated inverse gamma prior for $\sigma^2$, and a uniform prior (with support allowing small ranges up to ranges a
bit larger than the maximum distance over the region) for the decay parameter $3/a_e$. More specifically, the prior of $\mu$ is a Gaussian distribution with zero mean and variance one. For $\sigma^2$, we consider an inverse Gamma, ${\rm IG}(a_{\sigma},b_{\sigma})$, with $a_{\sigma}=2$ and $b_{\sigma}=1$, then the mean is $b_{\sigma}/(a_{\sigma}-1)=1$. The {continuous uniform distribution defined on the interval} $(0.001,1)$  is used {as the prior} for the decay parameter. The parameters are updated using a Metropolis--Hastings algorithm {\citep{hastings1970monte}}. For further details on the wrapped Gaussian spatial model fitting we refer to \citet{jona2012spatial}. The mean of the posteriori parameter estimates are considered in Step 3 of Algorithm \ref{alg_3}. Notice that in this case, {the} circular spatially correlated errors are generated  from wrapped Gaussian {spatial} processes, and in Step 2 of Algorithm \ref{alg_3}, a wrapped Gaussian spatial  process model is employed for model fitting. {This} modeling {only} allows  symmetric
marginal distributions. Therefore,  if the errors were drawn by using other {procedure}, such as {a} projected Gaussian {spatial process} (with asymmetric marginals), it would be more  convenient to use an alternative approach. 

In order to analyze the effect of the bandwidth matrix in the test statistics, $T_{n,p}^{1}$ and  $T_{n,p}^{2}$ are computed  in a grid of several bandwidths. {As in the experiment shown in Section \ref{sec:seve} in a independence framework,}  the bandwidth matrix is restricted to be diagonal with both equal elements,  $\mathbf{H}=\text{diag}(h,h)$. {In this case, the} different  values of $h=0.25, 0.40, 0.55, 0.70, 0.85, 1.00, 1.15, 1.30$ {are considered. Moreover, the same weight function $w$ as in Section \ref{sec:seve} is used here.}

Table \ref{table::c_simu_n_sp_ss} shows the {proportions} of rejections of the null hypothesis for different sample sizes and $\alpha=0.05$, when using $T_{n,p}^{1}$. Under the null hypothesis $(c=0)$,  it can be observed that the test has an acceptable performance using both bootstrap approaches PSCB and NPSCB. The {proportions} of rejections are similar to the theoretical level considered, namely $\alpha=0.05$. However, these proportions {clearly} depend on the value of the bandwidth $h$. For alternative assumptions $(c=1$ and $c=2)$, the performance of the test is satisfactory. As expected, the power of the test is larger when the value of $c$ is also larger. A slightly better performance of the is obtained when considering {the test statistic} $T_{n,p}^{2}$. {In this case, the proportions} of rejections of the null hypothesis are presented in Table \ref{table::c_simu_n_sp}.

Results for $n=400$ and different spatial dependence degrees ($a_e=0.1, 0.3,0.6$) are shown in Table \ref{table::c_simu_ae_sp_ss}, when using $T_{n,p}^{1}$.  PSCB and NPSCB approaches provide good results for {both} the null and the  alternative hypotheses. As expected, the power of the test is larger when the dependence structure is weaker. {In these scenarios,} results considering the test statistic $T_{n,p}^{2}$ are summarized in Table \ref{table::c_simu_ae_sp}.

\section{Conclusions and further research}\label{sec:conclusions_gof_c}
Testing procedures for assessing a parametric circular regression model (with circular response and $\mathbb{R}^d$-valued {covariate}) were proposed and analyzed in this work for independent and  for spatially correlated data. Specifically,  the test statistics were constructed by measuring a circular distance between a parametric fit ({non-smoothed} or smoothed) and a nonparametric estimator of the circular regression function. For the parametric approach, taking into account that the classical least squares regression method is not appropriate when the response variable is of circular nature, a circular analog  was used {\citep{fisher1992regression,lund1999least}}. Other parametric {fitting approaches}, such as maximum likelihood methods, could be used instead. Regarding the nonparametric fit, local polynomial type estimators were considered in the test {statistics}. 

Although the asymptotic distribution of the tests, under the null and under local alternatives, is out of the scope of this work, some guidelines to calculate this expression are provided in this section. As pointed out by \citet{kim2017multivariate}, using Taylor series expansions, the function $1-\cos (\Theta)$  can be approximated by $\Theta^2/2$, for $\Theta\in[0,2\pi)$. Therefore, the expressions $1-\cos[{\hat{m}}_{\mathbf{H}}(\mathbf{x};p)-{{m}}_{\hat{\bm{\beta}}}(\mathbf{x})]$ and $1-\cos[{\hat{m}}_{\mathbf{H}}(\mathbf{x};p)-{\hat{m}}_{\mathbf{H},\hat{\bm{\beta}}}(\mathbf{x};p)]$ in the test statistics $T_{n,p}^{1}$ and $T_{n,p}^{2}$, given in  (\ref{eq:statistic_circular_1}) and in (\ref{eq:statistic_circular_2}), respectively, can be approximated by $1/2[{\hat{m}}_{\mathbf{H}}(\mathbf{x};p)-{{m}}_{\hat{\bm{\beta}}}(\mathbf{x})]^2$ and $1/2[{\hat{m}}_{\mathbf{H}}(\mathbf{x};p)-{\hat{m}}_{\mathbf{H},\hat{\bm{\beta}}}(\mathbf{x};p)]^2$, respectively. Consequently,  $T_{n,p}^{1}$ and $T_{n,p}^{2}$ can be approximated by test statistics similar to {the ones used, for example, in \citet{hardle1993comparing} or in \citet{meilan2019goodness}}, for regression models with  Euclidean response and covariates. Notice that the  regression estimators involved in the test statistics $T_{n,p}^{1}$ and $T_{n,p}^{2}$ have more complicated expressions than those in {\citet{hardle1993comparing} or in \citet{meilan2019goodness}}. Therefore, as intuition suggests, it will be more difficult to calculate close {expressions} of their asymptotic {distributions}.

For practical implementation, bootstrap resampling methods were used to calibrate the test. For independent data, two procedures {have} been designed and compared: PCB and NPCB. Both methods are based on computing the residuals and generating independent bootstrap resamples. The main difference between them is the mechanism employed to obtain the residuals. In PCB, the residuals
come from the parametric regression  estimator. Alternatively, in NPCB, the residuals
are obtained from the nonparametric regression estimator.  For dependent data, in order to imitate the distribution of the (spatially {correlated}) errors, new bootstrap procedures were proposed: PSCB and NPSCB. Again, the main difference between both approaches is how the residuals are obtained. In the case of the PSCB, the residuals come  from the parametric fit, whereas in NPSCB, the residuals are obtained from the nonparametric estimator. In practice, in order to implement the procedure, a wrapped Gaussian spatial process model {\citep{jona2012spatial}} was fitted to them to mimic the dependence structure. {This} wrapped Gaussian spatial process model was fitted  within a Bayesian framework, therefore, some prior parameter values must be provided to use the Markov Chain Monte Carlo model fitting.  For further details on wrapped Gaussian model fitting, we refer to \citet{jona2012spatial}.  Alternatively,  other spatial-circular process models, such as asymmetric wrapped Gaussian spatial processes {\citep{mastrantonio2016wrapped}} or projected Gaussian spatial processes {\citep{wang2014modeling}} could be employed to model the residuals, and thus try to imitate the dependence structure of the errors. Notice that {once} the model is fitted, error bootstrap samples are generated from it. 
{These errors bootstrap samples could be also employed to design a parametric iterative least squares estimator, accounting for the possible spatial dependence structure, that could be used in the tests for spatially correlated data (instead of the parametric fit given in Section (\ref{eq:statistic_circular_1})). Specifically, using the errors bootstrap samples, the variance-covariance matrix of the circular errors can be approximated.   Then, applying a Cholesky decomposition of this matrix, the original circular responses and the $\mathbb{R}^d$-valued covariate are transformed, as it is done in the generalized least squares method. Finally, the parameter estimate is obtained applying (\ref{beta_circular_est_LS}) to the transformed observations. Obviously, this algorithm could be applied iteratively. Although, we have not applied this method in practice, we do not believe that it provides great improvements over using the circular least squares method described in Section \ref{sec:gof_circular_estimation_par}, even though the data are indeed dependent. The possible benefits of taking the correlation of the data into account  could be offset by the difficulty of adequately estimating the varaince-covariance matrix of the circular errors.}

For independent data, in the majority of scenarios considered in the simulation study, results obtained with NPCB improve those achieved by PCB, especially, for alternative assumptions. Moreover, a better behavior is observed when $T_{n,p}^{2}$, given in (\ref{eq:statistic_circular_2}), is employed. In general, the local linear type estimator seems to show a slightly better performance.  For spatially correlated {data}, it can be obtained that {both tests} do not  work properly under the null hypothesis, when using PCB and NPCB{ designed for independence}. Regarding  PSCB and NPSCB,  the use of nonparametric residuals in the bootstrap procedure provides the best results. As expected, the power of the test is larger when the spatial dependence structure is weaker. More satisfactory results are achieved when  $T_{n,1}^{2}$ is used.  In both frameworks (independent and spatially correlated data), the {proportions} of rejections of the null hypothesis {clearly depend} on the bandwidth matrix considered.

The procedures used in the simulation study  were implemented in
the statistical environment \texttt{R} \citep{Rsoft}, using functions included in the   \texttt{npsp} and \texttt{CircSpaceTime} packages  \citep{npsp, CircSpaceTime}.

\section*{Acknowledgements}\label{acknowledgements}
The authors acknowledge the support from the Xunta de Galicia grant ED481A-2017/361 and the European Union (European Social Fund - ESF). This research has been partially supported by MINECO grants  MTM2016-76969-P and MTM2017-82724-R, and by the Xunta de Galicia (Grupo de Referencia Competitiva  ED431C-2017-38, and Centro de Investigación del SUG ED431G 2019/01), all of them through the ERDF.

\begin{table}[H]
	\centering
	\resizebox{15cm}{!} {
\begin{tabular}{lclccccccc}
			Estimator&	$c$			&$n$&Method&$h=0.15$&$h=0.25$&$h=0.35$&$h=0.45$&$h=0.55$&$h=0.65$\\
			\hline
			Nadaraya---Watson	& 0 & 50 & PCB & 0.026 & 0.036 & 0.038 & 0.036 & 0.036 & 0.042 \\ 
			&  &  & NPCB & 0.008 & 0.008 & 0.012 & 0.018 & 0.020 & 0.022 \\ 
			&  & 100 & PCB & 0.024 & 0.030 & 0.026 & 0.028 & 0.028 & 0.030 \\ 
			&  &  & NPCB & 0.006 & 0.008 & 0.016 & 0.014 & 0.016 & 0.018 \\ 
			&  & 200 & PCB & 0.018 & 0.016 & 0.028 & 0.028 & 0.028 & 0.034 \\ 
			&  &  & NPCB & 0.000 & 0.002 & 0.008 & 0.010 & 0.008 & 0.008 \\ 
			\hline& 1 & 50 & PCB & 0.096 & 0.078 & 0.054 & 0.044 & 0.034 & 0.026 \\
\hline
			
	\end{tabular}}
	\caption{	Proportions of rejections of the null hypothesis for the parametric family  ${\mathcal{M}}_{1,\mathbf\beta}$ with different sample sizes and $\kappa=10$. The test statistic $T^{1}_{n,p}$ for $p=0,1$ is employed, and the Algorithm \ref{alg_2} is used. Significance level: $\alpha=0.05$.}
	\label{table::c_simu_n_m1_ss_uni}
\end{table}

\begin{table}[H]
	\centering
	\resizebox{15cm}{!} {
		\begin{tabular}{lclccccccc}
			Estimator&	$c$			&$n$&Method&$h=0.15$&$h=0.25$&$h=0.35$&$h=0.45$&$h=0.55$&$h=0.65$\\
			\hline
			Nadaraya--Watson	& 0 & 50 & PCB & 0.030 & 0.032 & 0.032 & 0.034 & 0.034 & 0.032 \\ 
			&  &  & NPCB & 0.008 & 0.010 & 0.010 & 0.012 & 0.014 & 0.014 \\ 
			&  & 100 & PCB & 0.022 & 0.028 & 0.024 & 0.028 & 0.028 & 0.030 \\ 
			&  &  & NPCB & 0.006 & 0.006 & 0.010 & 0.008 & 0.008 & 0.010 \\ 
			&  & 200 & PCB & 0.020 & 0.024 & 0.034 & 0.024 & 0.022 & 0.026 \\ 
			&  &  & NPCB & 0.004 & 0.004 & 0.004 & 0.004 & 0.004 & 0.004 \\ 
			\hline	& 1 & 50 & PCB & 0.092 & 0.102 & 0.098 & 0.090 & 0.070 & 0.066 \\
\hline

	\end{tabular}}
	\caption{	Proportions of rejections of the null hypothesis for the parametric family  ${\mathcal{M}}_{1,\mathbf\beta}$ with different sample sizes and $\kappa=10$. The test statistic $T^{2}_{n,p}$ for $p=0,1$ is employed, and the Algorithm \ref{alg_2} is used. Significance level: $\alpha=0.05$.}
	\label{table::c_simu_n_m1_uni}
\end{table}

\begin{table}[H]
	\centering
	\resizebox{15cm}{!} {
		\begin{tabular}{lclccccccc}
			Estimator&	$c$&			$\kappa$&Method&$h=0.15$&$h=0.25$&$h=0.35$&$h=0.45$&$h=0.55$&$h=0.65$\\
			\hline
			Nadaraya--Watson	& 0 & 5 & PCB & 0.034 & 0.024 & 0.032 & 0.030 & 0.040 & 0.044 \\ 
			&  &  & NPCB & 0.012 & 0.010 & 0.008 & 0.014 & 0.018 & 0.018 \\ 
			&  & 10 & PCB & 0.018 & 0.016 & 0.028 & 0.028 & 0.028 & 0.034 \\ 
			&  &  & NPCB & 0.000 & 0.002 & 0.008 & 0.010 & 0.008 & 0.008 \\ 
			&  & 15 & PCB & 0.028 & 0.022 & 0.032 & 0.032 & 0.026 & 0.026 \\ 
			&  &  & NPCB & 0.002 & 0.006 & 0.006 & 0.006 & 0.008 & 0.008 \\ 
			\hline & 1 & 5 & PCB & 0.148 & 0.088 & 0.048 & 0.024 & 0.014 & 0.008 \\
\hline

	\end{tabular}}
	\caption{	 Proportions of rejections of the null hypothesis for the parametric family  ${\mathcal{M}}_{1,\mathbf\beta}$ with different values of $\kappa$ and $n=400$. The test statistic $T^{1}_{n,p}$ for $p=0,1$ is employed, and the Algorithm \ref{alg_2} is used.  Significance level: $\alpha=0.05$.}
	\label{table::c_simu_k_m1_ss_uni}
\end{table}

\begin{table}[H]
	\centering
	\resizebox{15cm}{!} {
		\begin{tabular}{lclccccccc}
			Estimator&	$c$&			$\kappa$&Method&$h=0.15$&$h=0.25$&$h=0.35$&$h=0.45$&$h=0.55$&$h=0.65$\\
			\hline
			Nadaraya--Watson	& 0 & 5 & PCB & 0.034 & 0.026 & 0.024 & 0.030 & 0.026 & 0.028 \\ 
			&  &  & NPCB & 0.010 & 0.010 & 0.010 & 0.010 & 0.008 & 0.010 \\ 
			&  & 10 & PCB & 0.020 & 0.024 & 0.034 & 0.024 & 0.022 & 0.026 \\ 
			&  &  & NPCB & 0.004 & 0.004 & 0.004 & 0.004 & 0.004 & 0.004 \\ 
			&  & 15 & PCB & 0.024 & 0.034 & 0.038 & 0.028 & 0.028 & 0.026 \\ 
			&  &  & NPCB & 0.000 & 0.002 & 0.002 & 0.002 & 0.002 & 0.002 \\ 
			\hline	& 1 & 5 & PCB & 0.176 & 0.166 & 0.142 & 0.110 & 0.090 & 0.076 \\
\hline

	\end{tabular}}
	\caption{	 Proportions of rejections of the null hypothesis for the parametric family  ${\mathcal{M}}_{1,\mathbf\beta}$ with different values of $\kappa$ and $n=400$. The test statistic $T^{2}_{n,p}$ for $p=0,1$ is employed, and the Algorithm \ref{alg_2} is used. Significance level: $\alpha=0.05$.}
	\label{table::c_simu_k_m1_uni}
\end{table}

\begin{table}[H]
	\centering
	\resizebox{15cm}{!} {
		\begin{tabular}{lclcccccccc}
			Estimator&	$c$			&$n$&Method&$h=0.25$&$h=0.35$&$h=0.45$&$h=0.55$&$h=0.65$&$h=0.75$&$h=0.85$\\
			\hline
			Nadaraya--Watson& 0 & 100 & PCB & 0.030 & 0.038 & 0.042 & 0.048 & 0.062 & 0.072 & 0.072 \\ 
			&  &  & NPCB & 0.000 & 0.012 & 0.030 & 0.036 & 0.042 & 0.044 & 0.042 \\ 
			&  & 225 & PCB & 0.028 & 0.030 & 0.044 & 0.054 & 0.054 & 0.050 & 0.046 \\ 
			&  &  & NPCB & 0.002 & 0.008 & 0.016 & 0.028 & 0.026 & 0.024 & 0.022 \\ 
			&  & 400 & PCB & 0.038 & 0.044 & 0.036 & 0.034 & 0.034 & 0.032 & 0.026 \\ 
			&  &  & NPCB & 0.002 & 0.010 & 0.012 & 0.014 & 0.016 & 0.014 & 0.014 \\ 
			\hline	& 1 & 100 & PCB & 0.104 & 0.078 & 0.048 & 0.024 & 0.010 & 0.008 & 0.008 \\
\hline
	\end{tabular}}
	\caption{	Proportions of rejections of the null hypothesis for the parametric family  ${\mathcal{M}}_{2,\mathbf\beta}$ with different sample sizes and $\kappa=10$. The test statistic $T^{1}_{n,p}$ for $p=0,1$ is employed, and the Algorithm \ref{alg_2} is used. Significance level: $\alpha=0.05$.}
	\label{table::c_simu_n_m1_ss}
\end{table}

\begin{table}[H]
	\centering
	\resizebox{15cm}{!} {
		\begin{tabular}{lclcccccccc}
			Estimator&	$c$			&$n$&Method&$h=0.25$&$h=0.35$&$h=0.45$&$h=0.55$&$h=0.65$&$h=0.75$&$h=0.85$\\
			\hline
			Nadaraya--Watson& 0 & 100 & PCB & 0.032 & 0.036 & 0.044 & 0.046 & 0.050 & 0.052 & 0.052 \\ 
			&  &  & NPCB & 0.044 & 0.048 & 0.052 & 0.050 & 0.058 & 0.060 & 0.058 \\ 
			&  & 225 & PCB & 0.022 & 0.022 & 0.026 & 0.022 & 0.030 & 0.028 & 0.026 \\ 
			&  &  & NPCB & 0.026 & 0.026 & 0.022 & 0.022 & 0.022 & 0.026 & 0.024 \\ 
			&  & 400 & PCB & 0.030 & 0.034 & 0.040 & 0.042 & 0.036 & 0.038 & 0.030 \\ 
			&  &  & NPCB & 0.032 & 0.032 & 0.040 & 0.036 & 0.030 & 0.034 & 0.034 \\ 
			\hline&1 & 100 & PCB &  0.196 & 0.292 & 0.328 & 0.334 & 0.330 & 0.324 & 0.318 \\
\hline
	\end{tabular}}
	\caption{	Proportions of rejections of the null hypothesis for the parametric family  ${\mathcal{M}}_{2,\mathbf\beta}$ with different sample sizes and $\kappa=10$. The test statistic $T^{2}_{n,p}$ for $p=0,1$ is employed, and the Algorithm \ref{alg_2} is used. Significance level: $\alpha=0.05$.}
	\label{table::c_simu_n_m1}
\end{table}

\begin{table}[H]
	\centering
	\resizebox{15cm}{!} {
		\begin{tabular}{lclcccccccc}
			Estimator&	$c$&			$\kappa$&Method&$h=0.25$&$h=0.35$&$h=0.45$&$h=0.55$&$h=0.65$&$h=0.75$&$h=0.85$\\\hline
			Nadaraya--Watson & 0 & 5 & PCB & 0.030 & 0.030 & 0.042 & 0.042 & 0.042 & 0.036 & 0.034 \\ 
			&  &  & NPCB & 0.008 & 0.022 & 0.028 & 0.028 & 0.034 & 0.032 & 0.030 \\ 
			&  & 10 & PCB & 0.038 & 0.040 & 0.036 & 0.032 & 0.032 & 0.028 & 0.026 \\ 
			&  &  & NPCB & 0.002 & 0.010 & 0.012 & 0.014 & 0.014 & 0.016 & 0.014 \\ 
			&  & 15 & PCB & 0.040 & 0.044 & 0.042 & 0.034 & 0.034 & 0.032 & 0.028 \\ 
			&  &  & NPCB & 0.000 & 0.004 & 0.012 & 0.008 & 0.012 & 0.008 & 0.006 \\ 
			\hline	& 1 & 5 & PCB & 0.378 & 0.310 & 0.152 & 0.070 & 0.032 & 0.018 & 0.006 \\
\hline
			
	\end{tabular}}
	\caption{	 Proportions of rejections of the null hypothesis for the parametric family  ${\mathcal{M}}_{2,\mathbf\beta}$ with different values of $\kappa$ and $n=400$. The test statistic $T^{1}_{n,p}$ for $p=0,1$ is employed, and the Algorithm \ref{alg_2} is used.  Significance level: $\alpha=0.05$.}
	\label{table::c_simu_k_m1_ss}
\end{table}

\begin{table}[H]
	\centering
	\resizebox{15cm}{!} {
		\begin{tabular}{lclcccccccc}
			Estimator&	$c$&			$\kappa$&Method&$h=0.25$&$h=0.35$&$h=0.45$&$h=0.55$&$h=0.65$&$h=0.75$&$h=0.85$\\
			\hline
			Nadaraya--Watson& 0 & 5 & PCB & 0.026 & 0.028 & 0.036 & 0.030 & 0.030 & 0.026 & 0.028 \\ 
			&  &  & NPCB & 0.004 & 0.004 & 0.010 & 0.014 & 0.016 & 0.020 & 0.016 \\ 
			&  & 10 & PCB & 0.030 & 0.034 & 0.040 & 0.042 & 0.036 & 0.038 & 0.030 \\ 
			&  &  & NPCB & 0.032 & 0.032 & 0.040 & 0.036 & 0.030 & 0.034 & 0.034 \\ 
			&  & 15 & PCB & 0.032 & 0.034 & 0.042 & 0.044 & 0.036 & 0.036 & 0.036 \\ 
			&  &  & NPCB  & 0.024 & 0.024 & 0.038 & 0.044 & 0.042 & 0.036 & 0.036 \\  
			\hline	& 1& 5 & PCB & 0.492 & 0.544 & 0.542 & 0.538 & 0.522 & 0.504 & 0.490 \\
\hline
	\end{tabular}}
	\caption{	 Proportions of rejections of the null hypothesis for the parametric family  ${\mathcal{M}}_{2,\mathbf\beta}$ with different values of $\kappa$ and $n=400$. The test statistic $T^{2}_{n,p}$ for $p=0,1$ is employed, and the Algorithm \ref{alg_2} is used. Significance level: $\alpha=0.05$.}
	\label{table::c_simu_k_m1}
\end{table}

\begin{table}[H]
	\centering
	\resizebox{15cm}{!} {
		\begin{tabular}{lclccccccccc}
			Estimator&	$c$			&$n$&Method&$h=0.25$&$h=0.40$&$h=0.55$&$h=0.70$&$h=0.85$&$h=1.00$&$h=1.15$&$h=1.30$\\
			\hline
			Nadaraya--Watson	& 0 & 100 & PCB & 0.694 & 0.696 & 0.566 & 0.388 & 0.182 & 0.096 & 0.072 & 0.048 \\ 
			&  &  & NPCB & 0.224 & 0.380 & 0.336 & 0.232 & 0.108 & 0.056 & 0.032 & 0.026 \\ 
			&  & 225 & PCB & 0.994 & 0.978 & 0.902 & 0.608 & 0.348 & 0.190 & 0.108 & 0.084 \\ 
			&  &  & NPCB & 0.930 & 0.902 & 0.760 & 0.484 & 0.266 & 0.120 & 0.052 & 0.038 \\ 
			&  & 400 & PCB & 1.000 & 0.994 & 0.962 & 0.732 & 0.426 & 0.240 & 0.162 & 0.132 \\ 
			&  &  & NPCB & 1.000 & 0.982 & 0.914 & 0.632 & 0.348 & 0.172 & 0.106 & 0.084 \\ 
			\hline 			& 1 & 100 & PCB & 0.446 & 0.474 & 0.384 & 0.310 & 0.256 & 0.146 & 0.056 & 0.028 \\
\hline 			
			
	\end{tabular}}
	\caption{	Proportions of rejections of the null hypothesis for the parametric family  $\mathcal{M}_{1,\mathbf\beta}$ with different sample sizes. Model parameters:  $\sigma^2=0.16$ and $a_e=0.3$. The test statistic $T^{1}_{n,p}$ for $p=0,1$ is employed, and the Algorithm \ref{alg_2} is used. Significance level: $\alpha=0.05$.}
	\label{table::c_simu_n_sp_ind_ss}
\end{table}

\begin{table}[H]
	\centering
	\resizebox{15cm}{!} {
		\begin{tabular}{lclccccccccc}
			Estimator&	$c$			&$n$&Method&$h=0.25$&$h=0.40$&$h=0.55$&$h=0.70$&$h=0.85$&$h=1.00$&$h=1.15$&$h=1.30$\\
			\hline
			Nadaraya--Watson& 0 & 100 & PCB & 0.698 & 0.690 & 0.560 & 0.486 & 0.432 & 0.324 & 0.268 & 0.232 \\ 
			&  &  & NPCB & 0.218 & 0.362 & 0.332 & 0.288 & 0.228 & 0.170 & 0.144 & 0.126 \\ 
			&  & 225 & PCB & 0.996 & 0.982 & 0.928 & 0.856 & 0.780 & 0.682 & 0.580 & 0.530 \\ 
			&  &  & NPCB & 0.936 & 0.902 & 0.790 & 0.698 & 0.596 & 0.482 & 0.400 & 0.366 \\ 
			&  & 400 & PCB & 1.000 & 0.994 & 0.984 & 0.958 & 0.892 & 0.836 & 0.772 & 0.706 \\ 
			&  &  & NPCB & 1.000 & 0.986 & 0.942 & 0.868 & 0.780 & 0.700 & 0.598 & 0.548 \\ 
			\hline	 & 1 & 100 & PCB & 0.448 & 0.510 & 0.470 & 0.428 & 0.410 & 0.408 & 0.404 & 0.400 \\
\hline
			
	\end{tabular}}
	\caption{	Proportions of rejections of the null hypothesis for the parametric family  $\mathcal{M}_{1,\mathbf\beta}$ with different sample sizes. Model parameters:  $\sigma^2=0.16$ and $a_e=0.3$. The test statistic $T^{2}_{n,p}$ for $p=0,1$ is employed, and the Algorithm \ref{alg_2} is used. Significance level: $\alpha=0.05$.}
	\label{table::c_simu_n_sp_ind}
\end{table}

\begin{table}[H]
	\centering
	\resizebox{15cm}{!} {
		\begin{tabular}{lclccccccccc}
			Estimator&	$c$			&$a_e$&Method&$h=0.25$&$h=0.40$&$h=0.55$&$h=0.70$&$h=0.85$&$h=1.00$&$h=1.15$&$h=1.30$\\
			\hline Nadaraya--Watson
			& 0 & 0.1 & PCB & 0.864 & 0.624 & 0.336 & 0.102 & 0.042 & 0.016 & 0.006 & 0.004 \\ 
			&  &  & NPCB & 0.388 & 0.296 & 0.156 & 0.058 & 0.020 & 0.004 & 0.002 & 0.000 \\ 
			&  & 0.3 & PCB & 1.000 & 0.994 & 0.962 & 0.732 & 0.426 & 0.240 & 0.162 & 0.132 \\ 
			&  &  & NPCB & 1.000 & 0.982 & 0.914 & 0.632 & 0.348 & 0.172 & 0.106 & 0.084 \\ 
			&  & 0.6 & PCB & 0.986 & 0.976 & 0.948 & 0.870 & 0.654 & 0.432 & 0.296 & 0.246 \\ 
			&  &  & NPCB & 0.986 & 0.974 & 0.930 & 0.810 & 0.558 & 0.364 & 0.238 & 0.194 \\ 
			\hline	& 1 & 0.1 & PCB & 0.794 & 0.644 & 0.500 & 0.326 & 0.222 & 0.170 & 0.114 & 0.068 \\
\hline
			
	\end{tabular}}
	\caption{Proportions of rejections of the null hypothesis for the parametric family  ${\mathcal{M}}_{2,\mathbf\beta}$ with different values of $a_e$. Model parameters: $\sigma^2=0.16$ and $n=400$. The test statistic $T^{1}_{n,p}$ for $p=0,1$ is employed, and the Algorithm \ref{alg_2} is used. Significance level: $\alpha=0.05$.}
	\label{table::c_simu_ae_sp_ind_ss}
\end{table}

\begin{table}[H]
	\centering
	\resizebox{15cm}{!} {
		\begin{tabular}{lclccccccccc}
			Estimator&	$c$			&$a_e$&Method&$h=0.25$&$h=0.40$&$h=0.55$&$h=0.70$&$h=0.85$&$h=1.00$&$h=1.15$&$h=1.30$\\
			\hline
			Nadaraya--Watson	& 0 & 0.1 & PCB & 0.860 & 0.640 & 0.472 & 0.378 & 0.318 & 0.254 & 0.210 & 0.180 \\ 
			&  &  & NPCB & 0.388 & 0.294 & 0.218 & 0.168 & 0.130 & 0.104 & 0.074 & 0.062 \\ 
			&  & 0.3 & PCB & 1.000 & 0.994 & 0.984 & 0.958 & 0.892 & 0.836 & 0.772 & 0.706 \\ 
			&  &  & NPCB & 1.000 & 0.986 & 0.942 & 0.868 & 0.780 & 0.700 & 0.598 & 0.548 \\ 
			&  & 0.6 & PCB & 0.988 & 0.978 & 0.960 & 0.942 & 0.928 & 0.902 & 0.874 & 0.864 \\ 
			&  &  & NPCB & 0.988 & 0.978 & 0.958 & 0.928 & 0.876 & 0.830 & 0.790 & 0.756 \\ 
			\hline	& 1 & 0.1 & PCB & 0.826 & 0.782 & 0.754 & 0.730 & 0.712 & 0.716 & 0.714 & 0.716 \\
\hline
			
	\end{tabular}}
	\caption{Proportions of rejections of the null hypothesis for the parametric family  ${\mathcal{M}}_{2,\mathbf\beta}$ with different values of $a_e$. Model parameters: $\sigma^2=0.16$ and $n=400$. The test statistic $T^{2}_{n,p}$ for $p=0,1$ is employed, and the Algorithm \ref{alg_2} is used. Significance level: $\alpha=0.05$.}
	\label{table::c_simu_ae_sp_ind}
\end{table}

\begin{table}[H]
	\centering
	\resizebox{15cm}{!} {
		\begin{tabular}{lclccccccccc}
			Estimator&	$c$			&$n$&Method&$h=0.25$&$h=0.40$&$h=0.55$&$h=0.70$&$h=0.85$&$h=1.00$&$h=1.15$&$h=1.30$\\
			\hline
			Nadaraya--Watson	  & 0 & 100 & PSCB & 0.030 & 0.030 & 0.024 & 0.032 & 0.018 & 0.010 & 0.000 & 0.000 \\ 
			&  &  & NPSCB & 0.018 & 0.020 & 0.014 & 0.018 & 0.014 & 0.004 & 0.000 & 0.000 \\ 
			&  & 225 & PSCB & 0.036 & 0.034 & 0.034 & 0.024 & 0.020 & 0.016 & 0.012 & 0.008 \\ 
			&  &  & NPSCB & 0.022 & 0.026 & 0.018 & 0.012 & 0.002 & 0.002 & 0.002 & 0.002 \\ 
			&  & 400 & PSCB & 0.032 & 0.034 & 0.038 & 0.024 & 0.012 & 0.006 & 0.000 & 0.000 \\ 
			&  &  & NPSCB & 0.024 & 0.028 & 0.032 & 0.018 & 0.010 & 0.004 & 0.000 & 0.000 \\ 
			
			\hline	& 1 & 100 & PSCB & 0.086 & 0.144 & 0.136 & 0.132 & 0.092 & 0.030 & 0.014 & 0.010 \\
\hline
			
	\end{tabular}}
	\caption{	Proportions of rejections of the null hypothesis for the parametric family  $\mathcal{M}_{1,\mathbf\beta}$ with different sample sizes. Model parameters:  $\sigma^2=0.16$ and $a_e=0.3$. The test statistic $T^{1}_{n,p}$ for $p=0,1$ is employed, and the Algorithm \ref{alg_3} is used. Significance level: $\alpha=0.05$.}
	\label{table::c_simu_n_sp_ss}
\end{table}

\begin{table}[H]
	\centering
	\resizebox{15cm}{!} {
		\begin{tabular}{lclccccccccc}
			Estimator&	$c$			&$n$&Method&$h=0.25$&$h=0.40$&$h=0.55$&$h=0.70$&$h=0.85$&$h=1.00$&$h=1.15$&$h=1.30$\\
			\hline
			Nadaraya--Watson& 0 & 100 & PSCB & 0.030 & 0.032 & 0.026 & 0.024 & 0.036 & 0.028 & 0.024 & 0.024 \\ 
			&  &  & NPSCB & 0.018 & 0.018 & 0.018 & 0.018 & 0.012 & 0.012 & 0.014 & 0.012 \\ 
			&  & 225 & PSCB & 0.036 & 0.034 & 0.034 & 0.024 & 0.020 & 0.016 & 0.012 & 0.008 \\ 
			&  &  & NPSCB & 0.022 & 0.022 & 0.024 & 0.018 & 0.014 & 0.006 & 0.004 & 0.006 \\ 
			&  & 400 & PSCB & 0.030 & 0.032 & 0.036 & 0.030 & 0.026 & 0.026 & 0.024 & 0.030 \\ 
			&  &  & NPSCB & 0.024 & 0.028 & 0.032 & 0.024 & 0.022 & 0.016 & 0.020 & 0.020 \\ 
			\hline	& 1 & 100 & PSCB & 0.086 & 0.154 & 0.176 & 0.194 & 0.198 & 0.200 & 0.192 & 0.178 \\
\hline
			
	\end{tabular}}
	\caption{	Proportions of rejections of the null hypothesis for the parametric family  $\mathcal{M}_{1,\mathbf\beta}$ with different sample sizes. Model parameters:  $\sigma^2=0.16$ and $a_e=0.3$. The test statistic $T^{2}_{n,p}$ for $p=0,1$ is employed, and the Algorithm \ref{alg_3} is used. Significance level: $\alpha=0.05$.}
	\label{table::c_simu_n_sp}
\end{table}

\begin{table}[H]
	\centering
	\resizebox{15cm}{!} {
		\begin{tabular}{lclccccccccc}
			Estimator&	$c$			&$n$&Method&$h=0.25$&$h=0.40$&$h=0.55$&$h=0.70$&$h=0.85$&$h=1.00$&$h=1.15$&$h=1.30$\\
			\hline
			Nadaraya--Watson& 0 & 0.1 & PSCB & 0.038 & 0.042 & 0.030 & 0.028 & 0.014 & 0.006 & 0.002 & 0.000 \\ 
			&  &  & NPSCB & 0.010 & 0.022 & 0.022 & 0.016 & 0.008 & 0.004 & 0.000 & 0.000 \\ 
			&  & 0.3 & PSCB & 0.032 & 0.034 & 0.038 & 0.024 & 0.012 & 0.006 & 0.000 & 0.000 \\ 
			&  &  & NPSCB & 0.024 & 0.028 & 0.032 & 0.018 & 0.010 & 0.004 & 0.000 & 0.000 \\ 
			&  & 0.6 & PSCB & 0.032 & 0.034 & 0.028 & 0.018 & 0.008 & 0.006 & 0.004 & 0.004 \\ 
			&  &  & NPSCB & 0.030 & 0.030 & 0.022 & 0.010 & 0.004 & 0.000 & 0.000 & 0.000 \\ 
			\hline& 1 & 0.1 & PSCB & 0.654 & 0.650 & 0.636 & 0.600 & 0.572 & 0.554 & 0.540 & 0.528 \\
\hline

	\end{tabular}}
	\caption{Proportions of rejections of the null hypothesis for the parametric family  ${\mathcal{M}}_{2,\mathbf\beta}$ with different values of $a_e$. Model parameters: $\sigma^2=0.16$ and $n=400$. The test statistic $T^{1}_{n,p}$ for $p=0,1$ is employed, and the Algorithm \ref{alg_3} is used. Significance level: $\alpha=0.05$.}
	\label{table::c_simu_ae_sp_ss}
\end{table}

\begin{table}[H]
	\centering
	\resizebox{15cm}{!} {
		\begin{tabular}{lclccccccccc}
			Estimator&	$c$			&$n$&Method&$h=0.25$&$h=0.40$&$h=0.55$&$h=0.70$&$h=0.85$&$h=1.00$&$h=1.15$&$h=1.30$\\
			\hline
			Nadaraya--Watson& 0 & 0.1 & PSCB & 0.038 & 0.038 & 0.038 & 0.038 & 0.042 & 0.042 & 0.042 & 0.044 \\ 
			&  &  & NPSCB & 0.010 & 0.020 & 0.032 & 0.028 & 0.022 & 0.024 & 0.034 & 0.028 \\ 
			&  & 0.3 & PSCB & 0.030 & 0.032 & 0.036 & 0.030 & 0.026 & 0.026 & 0.024 & 0.030 \\ 
			&  &  & NPSCB & 0.024 & 0.028 & 0.032 & 0.024 & 0.022 & 0.016 & 0.020 & 0.020 \\ 
			&  & 0.6 & PSCB & 0.030 & 0.030 & 0.030 & 0.018 & 0.014 & 0.006 & 0.006 & 0.006 \\ 
			&  &  & NPSCB & 0.030 & 0.024 & 0.028 & 0.014 & 0.008 & 0.004 & 0.004 & 0.004 \\ 
			\hline& 1 & 0.1 & PSCB & 0.640 & 0.648 & 0.634 & 0.620 & 0.606 & 0.598 & 0.598 & 0.596 \\
\hline
			
	\end{tabular}}
	\caption{Proportions of rejections of the null hypothesis for the parametric family  ${\mathcal{M}}_{2,\mathbf\beta}$ with different values of $a_e$. Model parameters: $\sigma^2=0.16$ and $n=400$. The test statistic $T^{2}_{n,p}$ for $p=0,1$ is employed, and the Algorithm \ref{alg_3} is used. Significance level: $\alpha=0.05$.}
	\label{table::c_simu_ae_sp}
\end{table}

\bibliographystyle{spbasic}   
\bibliography{bibibnew}

\end{document}